# On the Carroll-Chen Model

ABSTRACT

I argue that the Carroll-Chen cosmogenic model does not provide a plausible scientific explanation of our universe's initial low-entropy state.

**1** *Introduction*
**2** *The Carroll-Chen Model*
**3** *Inconsistency, Ambiguity, and Admitted Incompleteness*
**4** *Unbounded Entropy?*
**5** *Nucleation and Metagalaxy Creation*
**6** *Causation and the Carroll-Chen Model*
**7** *Conclusion*

## 1  Introduction

Our universe's initial low-entropy state could be explained.¹ In fact, some have tried to explain it.² Jonathan Schaffer's account of grounding and priority monism forces the conclusion that states of the cosmos are *metaphysically explained* by the cosmos itself.³ Axiarchic approaches to *principled-based explanation* suggest that some evaluative principle explains all of reality, including the initial low-entropy state.⁴ Quentin Smith ([2007], pp. 188-97) argued that every state of the cosmos is *causally explained* by some prior state, and that it is by virtue of such state-by-state explanation that the entire cosmos is accounted for, including the low-entropy state.⁵ There are also a great many attempts to *scientifically explain* the relevant state by appeal to inflation, pre-big bang models, and other developments in cosmology and cosmogeny.[6] Here I address just one of these potential scientific explanations, specifically the explanation proposed by a cosmogenic model recently developed by Sean Carroll and Jennifer Chen (hereafter I will refer to this model with the designation 'CC-M').

---

¹ Even some of those who would insist that such a state is brute believe that it could be explained. See (Callendar, [2004a], p. 199, though *cf.* his comments in [2004b], p. 241).

² It is worth emphasizing here, that most cosmologists working on the low-entropy initial condition vie for a dynamical explanation of that condition. As Andreas Albrecht ([2004], p. 374-5) noted, "…most cosmologists would instinctively take a different perspective. They would try and look further into the past and ask how such strange 'initial' conditions could possibly have been set up by whatever dynamical process went before."

³ This is because, on Schaffer's view, substances ground states, and because the cosmos grounds every other substance and also its own states. See (Schaffer [2009a], [2009b], [2010a], [2010b], [2013]). I am assuming that obtaining grounding relations (relations of dependence for nature and positive ontological status that are transitive, asymmetric, and well-founded) usually underwrite and back metaphysical explanations, much like obtaining causal relations usually underwrite and back causal explanations (see Woodward [2003], pp. 209-20. He seems to go in for the "backing" idea with respect to singular causal explanation. In fact, Strevens [2007, p. 237] remarked, "[t]he core of Woodward's account of singular event explanation is the account of singular event causation…")

⁴ I have in mind the work of Leslie ([1979]), Parfit ([2011], pp. 623-48), and Rescher ([2010]).

⁵ That is to say, the whole or entire collection of states is explained as soon as all of the parts or members of the collection are explained. See (Hume [1947]; and Paul Edwards [1959]), though the idea goes back to (William of Ockham [1957]; *cf.* Rescher's discussion in [2010], pp. 22-25).

My examination of the CC-M will proceed as follows: Sect. 2 provides an informal explication of the CC-M. Sect. 3 argues that the CC-M is internally inconsistent, ambiguously described, and admittedly incomplete. I suggest that the aforementioned inconsistency, ambiguity, and incompleteness implies that the recommended scientific explanation of our low-entropy state—a key motivation for proposing the CC-M—is not an actual scientific explanation of that state. Moreover, I argue in Sect. 4, that Carroll and Chen (henceforth C&C) cannot plausibly maintain that entropy is unbounded from above. Sect. 5 attempts to knock down the model's recommended mechanisms for universe nucleation out of a background empty or asymptotically de Sitter space-time, and I conclude the paper in sect. 6 with an objection to the model from the formal nature of the causal relation.

## 2   The Carroll-Chen Model

Our universe began in an extremely smooth, non-empty homogeneous state.[7] That initial non-empty smoothness or homogeneity just is the initial low-entropy state of the cosmos.[8] Our best science suggests that our arrow of time points in the direction of entropic increase, since our best science suggests that time's arrow *reduces* to the arrow of entropic increase.[9] C&C find

---

[6] On the explanation from inflation see (Davies [1983]; but more recently Guth [2004], p. 37). Though there are many, attempted pre-big bang explanations can be found in work on ekpyrotic and cyclic universe models, for which see (Khoury et. al. [2001]; Khoury, et. al. [2002]; and Steinhardt and Turok [2002a], [2002b], [2005]). Holographic cosmogenic models also attempt to explain the low-entropy state. See particularly (Banks [2007]).

[7] See on this (Davies [1980], pp. 160-1, and pp. 168-9; and Penrose [1981], pp. 247-9, [1989b]). According to observations which involve looking out past 300 million light years, the universe is generally homogenous and isotropic. See (Wu, Lahav, and Rees [1999], p. 225; and Weinberg [2008], p. 1).

The standard Freidman-Lemaître-Robertson-Walker (FLRW) model predicts that the universe is remarkably isotropic and homogenous (in the sense specified by Wald [1984], pp. 92-3). See also (Lyth and Liddle [2009], p. 39; Penrose [2005], p. 718), *inter alios*. I should add that the EGS theorem (proven in Ehlers, Geren, and Sachs [1968]) establishes that our universe is FLRW solely on the basis of isotropy in propagating cosmic microwave background radiation (CMBR), and the Copernican principle (Smeenk [2013], p. 631). The fact that our universe is FLRW suggests that the distribution of matter in our universe is uniform (Smeenk [2013, p. 613]). *Q.v.* sect. 5.1 for more on this and related results.

[8] The "early spatial uniformity represents the universe's extraordinarily low initial entropy" (Penrose [2010], p. 76; Penrose [2005], pp. 706-7). How low and high entropy states are to be understood when gravity is in play is somewhat controversial (see the comments in Egan and Lineweaver ([2010], p. 1826). Carroll, for example, takes issue with certain characterizations of maximum entropy. He believes that even black holes can increase in entropy by radiating away into empty space (hence my description of the initial low-entropy state as *non-empty* initial smoothness). See (Carroll [2010], pp. 302-3; Aguirre, Carroll, and Johnson [2011], pp. 18-9), and see (Page [2005], p. 10) for some precise details regarding black hole radiation emission and entropy increase. See also the broader discussions of entropy in (Albrecht [2004], pp. 371-4; Greene [2004], pp. 171-5; North [2011], p. 327; Penrose [1979], pp. 611-17, [1989b], pp. 251-7, [2010], pp. 73-9; Price [1996], pp. 79-83, [2004], pp. 227-8; and Wald [1984], pp. 416-8, [2006], p. 395). Callendar ([2010], pp. 47-51) articulates some problems for the standard way of understanding entropy and gravity. Earman ([2006], pp. 417-8, *cf*. the comments on p. 427) is very skeptical of the contemporary orthodoxy on these matters.

[9] Let me say here what I'm concerned with when I discuss or mention *the* arrow of time. First, I am not interested in the asymmetry of time itself. I am, however, concerned with the asymmetry of the contents of the cosmos (on this distinction see Price [1996], pp. 16-7; North [2011], p. 312). There are, therefore, many arrows of

these facts to be "unnatural".[10] Their model attempts to advance a promising strategy for understanding the arrow of time and initial smoothness naturally. The strategy itself recommends a scientific explanation of the initial smoothness and so also the arrow of time. This explanation has need of the conjecture that the initial low-entropy state was produced by way of "dynamical evolution from a generic state."[11] The following theses are indispensable to the proposed scientific explanation:

> (1): Our metagalaxy was produced by a background Universe which is an empty/pure (dS) or asymptotic (AsDS) de Sitter space-time.[12]
> (2): The Universe produced our metagalaxy by means of a fluctuation. Such a fluctuation gave birth to a proto-inflationary region. It was this region which sparked the process of eternal inflation that is responsible for the large-scale structure of our metagalaxy.[13]
> (3): Entropy is unbounded from above.

I will now informally discuss each claim, and in the process shed more light on less central aspects of the CC-M.

---

time, though some maintain that these arrows can be reduced to the thermodynamic arrow. It is this supposed principal arrow with which I'm worried when I comment on *the* arrow of time below.

    That the thermodynamic arrow of time should be understood in terms of the direction of entropic increase is the majority view these days. C&C ([2004], p. 3) remarked, "…the thermodynamic arrow of time is the direction picked out by this increase of entropy." Dyson, Kleban, and Suskind ([2002], p. 1) stated, "[t]he low entropy starting point is the ultimate reason that the universe has an arrow of time, without which the second law would not make sense." *Cf.* (Bousso [2012], pp. 2-3, and pp. 26-9) for a different view. The discussion of these sorts of issues in (North [2011]) is first-rate.

[10] Carroll declared that "[a]mong the unnatural aspects of the universe, one stands out: time asymmetry" Carroll ([2008a], p. 48). In ([ibid.], p. 50) he remarked, "[t]he question remains: Why was the entropy low to start with? It seems very unnatural given that low-entropy states are so rare". In his ([2006], p. 1132) he stated,

> "…the Universe that we observe seems remarkably unnatural. The entropy of the Universe is not nearly as large as it could be, although it is at least increasing; for some reason, the early Universe was in a state of incredibly low entropy."

    The notions of naturalness and unnaturalness are left at an intuitive level, though C&C do seem to connect unnaturalness with improbability at times (see Carroll and Chen [2004], p. 3). C&C's understanding of unnaturalness also clearly motivates their rejection of the doctrine that the initial smoothness of our cosmos is in no need of explanation. Carroll ([2006], p. 1132) stated, "[w]hen we come across a situation that seems unnatural or finely tuned, physicists seize upon it as a clue pointing towards some underlying mechanism that made it that way." *Cf*. (Carroll [2010], p. 288).

[11] (Carroll and Chen [2004], p. 6). They would go on to admit that, "[b]y taking seriously the ability of spacetime to expand and dilute degrees of freedom, we claim to have shown how an arrow of time can *naturally* arise dynamically in the course of the evolution from a generic boundary condition." (Carroll and Chen [2004], p. 29 emphasis mine). That the model portends to explain the low-entropy state is clear: "[t]his scenario explains why a universe like ours is likely to have begun via a period of inflation, and also provides an origin for the cosmological arrow of time" from (Carroll and Chen [2005], p. 1671).

[12] Below, I follow the convention of Russian cosmologists in regarding the universes that help comprise the multiverse as metagalaxies which are spawned somehow by a background space-time which I will (not necessarily following the convention of others) call the 'Universe' (capital-U). See (Glushkov [2005], p. 16 who seems to follow the former convention), and Leslie's ([1989], p. 1) point regarding the convention tied to the term 'metagalaxy'.

[13] See (Carroll and Chen [2004], p. 5).

2.1 The Background de Sitter Space and Unbounded Entropy

C&C seek a scientific explanation of our metagalaxy's initial low-entropy state that does not include finely-tuned boundary conditions or temporally asymmetric micro-dynamics.[14] In order to acquire such an explanation, C&C need a background Universe.[15] This background space-time, has an initial Cauchy hypersurface with generic conditions that are wholly natural. There is also a sense in which the entire background space is admitted to be natural. For C&C, however, "natural means high-entropy"[16], thus the background space-time can be understood as a "middle moment" (to borrow Carroll's wording) with the highest amount of entropy that an individual interrelated cosmos with a positive vacuum energy can have. Carroll wrote:

> That middle moment was not finely tuned to some special very-low-entropy initial condition, as in typical bouncing models. It was as high as we could get, for a single connected universe in the presence of a positive vacuum energy. That's the trick: allowing entropy to continue to rise in both directions of time, even though it started out large to begin with.[17]

In their ([2004]) depiction of the CC-M, the background space-time evolves in two directions away from some arbitrary generic initial surface. There is then further evolution on both sides of the surface into de Sitter phases with a positive cosmological constant.[18] Details about the nature of the initial surface are left to the imagination, though C&C suggest that such specifics are irrelevant. Allowing for evolution away from the initial surface in two directions implies that C&C do permit an understanding of the initial surface as a surface which constitutes the place over which a type of initial condition can be defined. That initial condition will not be "an equilibrium state with maximal entropy."[19] In fact, such a condition over the initial Cauchy surface will be the surface "of minimum entropy."[20] Thus, entropy increases away from the initial surface in two directions. Such dual entropic increase constitutes the dependency base for two arrows of time. As the two sides of space-time approach their respective de Sitter phases, each arrow of time will become in some sense ambiguous. This is because empty de Sitter phases are in thermal equilibrium states.[21] There is, therefore, no entropic increase once either side of the ultra-large scale structure reaches respective de Sitter phases, and this further implies that there are no arrows of time during the respective phases of the cosmic evolution of the Universe.

---

[14] See (Carroll and Chen [2004], p. 6 and p. 27).

[15] Carroll ([2008a], p. 48) stated,
> "[i]ncreasingly, however, this puzzle [of the arrow of time and entropy] about the universe we observe hints at the existence of a much larger spacetime we do not observe. It adds support to the notion that we are part of a multiverse whose dynamics help to explain the seemingly unnatural features of our local vicinity."

[16] (Carroll and Chen [2004], p. 7).

[17] (Carroll [2010], p. 362).

[18] See (Carroll and Chen [2004], pp. 28-29).

[19] (Carroll and Chen [2004], p. 27). I'm borrowing their wording here. The quotation in context is about something different, *viz.*, the fact that the background space is never in an equilibrium state because baby universes can always be generated resulting in the further increase of entropy.

[20] (Carroll and Chen [2004], p. 5).

[21] Carroll ([2010], p. 355 emphasis mine) said, "De Sitter space is empty apart from the thin background of thermal radiation, so for the most part it is completely inhospitable to life; *there is no arrow of time*, since it's in thermal equilibrium."

In subsequent work, Carroll seems to modify the CC-M (this modified version of the account will be individuated *via* the locution '$^M$CC-M').[22] $^M$CC-M's background space shares some affinities with the space-time described by Willem de Sitter's solution to Einstein's field equations. That solution's line element is as follows (using de Sitter's coordinates):

$$(4): ds^2 = -dr^2 - R^2 \sin^2\left(\frac{r}{R}\right)(d\psi^2 + \sin^2\psi \, d\theta^2) + \cos^2\left(\frac{r}{R}\right) c^2 \, dt^2 \quad \text{(Eq. 1)}^{23}$$

(Eq.1) predicts that matter (what de Sitter called "world-matter") is completely missing from the space, and so de Sitter's space is empty.[24] The background space of the $^M$CC-M is likewise empty.[25] (Eq. 1) suggests a metric which features a cosmological constant that is positive (*q.v.* note 23). In contemporary cosmology and astrophysics, the cosmological constant is thought to correspond to (dark) vacuum energy.[26] Thus, de Sitter's space-time includes a positive vacuum energy, and the same turns out to be true of the $^M$CC-M background space. Lastly, the geometry recommended by (Eq. 1) is such that the space described is hyperbolical.[27] More generally, de Sitter space-time is represented as a Lorentzian 4-sphere within a Minkowskian 5-space with the following metric $ds^2 = dt^2 - dw^2 - dx^2 - dy^2 - dz^2$.[28] (See the nice illustration of the space in Carroll [2006], p. 1135.)

Because the Universe on the $^M$CC-M is a pure de Sitter space-time, it is past-geodesically complete.[29] Moreover, Carroll describes the CC-M as one which starts with the assumption that

---

[22] As early as Carroll's ([2006], p. 1134), the background space-time seems to become empty de Sitter. One also gets a hint of this in (Carroll [2010], pp. 362-3).

[23] Given that $r_0 = 0$ and that $\lambda = \frac{3}{R^2}$; where R corresponds to a positive constant, and *r* is the Schwarzschild radius. The equation is from (de Sitter [1918], p. 230); but see also the discussion in (de Sitter [1917], p. 7; and Earman [1995], p. 7).

[24] (de Sitter [1918], p. 229). There was some early debate about whether or not de Sitter's space was truly empty. Einstein ([1918], p. 272) argued that the space contained singularities. Arthur Eddington ([1923], p. 165) would go on to correctly judge that the supposed singularities were merely coordinate. See the discussion of these matters in (Earman [1995], pp. 5-11).

[25] For all intents and purposes, the space is empty. Carroll ([2010], p. 355) remarked, "De Sitter space is empty apart from the thin background of thermal radiation…".

[26] Carroll ([2010], p. 308) defines vacuum energy as "a constant amount of energy inherent in every cubic centimeter of space, one that remains fixed throughout space and time."

It is interesting that some models try to get along without dark energy (see, for example, work on the Cardassian model in Freese [2003], p. 53; and Freese and Lewis [2002], p. 6).

[27] (de Sitter [1918], p. 233).

[28] (Penrose [2005], p. 747-8; Misner, Thorne, and Wheeler [1973], p. 745; and for an extensive treatment of de Sitter and anti-de Sitter space-times see Hawking and Ellis [1973], pp. 124-34; but *vide etiam* the discussions in Bousso [1998], [2000a]; and Ginsparg and Perry [1983], pp. 245-251). I should add here that de Sitter space is also thought to have infinite volume. See (Carroll and Chen [2004], p. 27).

[29] Hawking and Ellis ([1973], p. 126) remarked,
> "de Sitter space is geodesically complete; however, there are points in space which cannot be joined to each other by any geodesic. This is in contrast to spaces with a positive definite metric, when geodesic completeness guarantees that any two points of a space can be joined by at least one geodesic."

*Cf.* Bousso, DeWolfe, and Myers ([2003], p. 300), who stated, "…de Sitter space has two disconnected infinities, one in the past and one in the future".

the Universe is eternal,[30] and interpreters of his work commonly understand the model to be committed to such eternality as well.[31]

2.2 Nucleated Metagalaxies and Unbounded Entropy

de Sitter space is exceedingly cold, less than $10^{-28}$ °Kelvin, though its temperature is still above zero.[32] The temperature of de Sitter space-time is positive because it possesses (quoting Gibbons and Hawking) "thermal radiation with a characteristic wavelength of the order of the Hubble radius."[33] The fact that de Sitter space-time has a positive temperature implies that that space-time countenances fluctuations which result (it is hoped by C&C) in the existence of "…new inflating patches, which can eventually evolve into universes like ours".[34] With a positive vacuum energy, and the positive temperature of the background space, fluctuations can cause an inflaton field to ascend its potential so as to produce the beginning stages of eternal inflation (*i.e.*, the production of a sufficiently ample vacuum energy).[35] And while it is true that our metagalaxy began in a very low-entropy state, that state exhibited more entropy than the relevant "tiny commoving volume of de Sitter" space "from which it arose…"[36] This is because the entropy density *per* that tiny volume of de Sitter space is considerably low.[37] The fluctuations in de Sitter space are not random, but are the consequence of the obtaining of a certain condition that is itself produced by the space. C&C remarked, "[b]ecause the entropy *density* of the

---

[30] (Carroll [2010, p. 350], pp. 361-2). Elsewhere, Carroll and Chen ([2004], p. 5 emphasis mine) stated:
> "We also predict that this structure should be recovered *infinitely far into the past*, with a reversed thermodynamic arrow of time. Our overall universe is therefore statistically time-symmetric about some Cauchy surface of minimum entropy."

Carroll is forthright about his commitment to an eternal past. He appears to believe that a truly quantum cosmology will rub out our metagalaxy's initial singularity and put to rest the claim that that singularity was a veritable beginning of time (see Carroll [2008b], p. 4, *cf.* pp. 6-7).

[31] See (McInnes [2007], p. 22).

[32] (Carroll ([2010], p. 313; Gibbons and Hawking [1977], p. 2739).

[33] (Gibbons and Hawking [1977], p. 2739).

[34] (Carroll and Chen [2005], p. 1673). "In the presence of an appropriate inflaton field, thermal fluctuations will occasionally conspire to produce a tiny, smooth region of space dominated by a large vacuum energy—the correct conditions for a proto-inflationary patch [15]" from ([ibid.]).

[35] (Carroll and Chen [2004, p. 27]; Carroll [2006], p. 1133, [2008b], p. 8). With respect to how this might all precisely work, Carroll seems to rely heavily upon the tunneling story written down by Farhi, Guth, and Guven ([1990]), he remarked:
> "…de Sitter space, the solution of Einstein's equation in the presence of a positive cosmological constant, is unstable; there must be some way for it to undergo a transition into a state with even more entropy. *Chen and I imagined that the mechanism was the quantum creation of baby universes, as suggested by Farhi, Guth, and Guven* [14]" (Carroll [2008b], p. 8 emphasis mine. Note [14] in the text refers the reader to Farhi, Guth, and Guven [1990]).

There are some attempted improvements on the FGG model of nucleation. See, for example, (Fischler, Morgan, and Polchinski [1990]; and Linde [1992]). For a barrage of criticisms, see (Banks [2003]; Freivogel et. al. [2006]; and *cf.* Aguirre, Gratton, and Johnson [2007], p. 9).

[36] (Carroll and Chen [2004], p. 26).

[37] (Carroll and Chen [2005], p. 1673; Carroll [2006], p. 1133). This part of the story is significantly changed in Carroll's more recent and more technical work with Anthony Aguirre and Matthew C. Johnson. They argue that "cosmological fluctuations to lower-entropy states should be thought of as the *time reverse* of a—generally smooth, or at least gradual—natural evolution from a low-entropy state into equilibrium." (Aguirre, Carroll and Johnson [2011], p. 1 emphasis in the original).

background is so low, it is easier to fluctuate into a small proto-inflationary patch than into a universe that looks like ours today."[38] Thus, thermal fluctuations, in an almost completely empty de Sitter space in which there is low entropy density in the background, yield a proto-inflationary patch out of which our metagalaxy can form *via* the mechanism of eternal inflation.

Because advanced stages of the Universe's evolution are empty de Sitter on both the CC-M and $^M$CC-M, metagalaxy nucleation conditions arise. The birth of metagalaxy's with respective eternally inflating phases yields an avenue for unbounded entropic increase.[39]

The thesis of unbounded entropy has very clear implications. First, if (3) is true, then the amount of energy in the background space is infinite. Second, given (3), there are infinitely many degrees of freedom. And third, (3) implies that with respect to the Universe, there is no such thing as an entropic or thermodynamic equilibrium state. If any of these implications are proven false, it will follow by *modus tollens* that (3) is false as well.

## 3 Inconsistency, Ambiguity, and Admitted Incompleteness

In this section, I will seek to maintain the following contentions: First, articulations of the model are inconsistent.[40] Second, the CC-M is ambiguously described. Third, the scientific explanation of our initial non-empty and smooth state provided by the CC-M is *admittedly* incomplete.[41] And fourth, given such admitted incompleteness, the aforementioned explanation fails to account for our arrow of time.

On the CC-M, our metagalaxy is a closed and "essentially *autonomous*" system, "free from outside influences".[42] One might wonder how our metagalaxy achieved such independence on the CC-M. According to some of Carroll's work, such independence was achieved by means of the mechanism of metagalaxy nucleation developed by Edward Farhi, Alan Guth, and Jemal Guven ([1990], henceforth 'FGG'). On the FGG, when there is successful nucleation, metagalaxies completely separate from their mother Universe. Here is Carroll's description of the process:

---

[38] (Carroll and Chen [2005], p. 1673 emphasis in the original).

[39] Carroll ([2010], p. 360-1) admitted:
> "Once baby universes are added to the game, the system is no longer in equilibrium, for the simple reason that there is no such thing as equilibrium. In the presence of a positive vacuum energy (according to this story), the entropy of the universe never reaches a maximum value and stays there, because there is no maximum value for the entropy of the universe—it can always increase, by creating new universes….By suggesting that there is no such thing as equilibrium, we can avoid this dilemma. It becomes natural to observe entropy increasing, simply because entropy can always increase."

(*cf.* his remarks in [ibid.], p. 365)

[40] Unless otherwise indicated, in this section just about everything I say about the CC-M holds for the $^M$CC-M. Therefore, (again, unless I indicate otherwise) wherever one sees 'CC-M', read '$^M$CC-M' as well.

[41] Let me provide a bit of an *apologia* for what I'm up to in this section. First, both Carroll and Chen are completely honest and humble about the CC-M's incompleteness. I do not mean to mercilessly pile on their worries about how to complete the model. My contention below will be that given scientific realism and the fact that substantive portions of the CC-M are admittedly not well-understood, one cannot plausibly maintain that the CC-M provides a *bona fide* explanation of the low-entropy state. That is an important academic and philosophical point. Second, subsequent sections of this paper criticize the model on the assumption that there *are* ways of providing the details. So even if one does not agree with the aforementioned contention, one will still have to respond to some damaging criticism.

[42] (Carroll [2010], p. 335 emphasis in the original).

> What we see is simultaneous fluctuation of the inflaton field, creating a bubble of false vacuum, and of space itself, *creating a region that pinches off from the rest of the universe*. The tiny throat that connects the two is a wormhole…But this wormhole is unstable and will quickly collapse to nothing, leaving us with *two disconnected spacetimes*: the original parent universe and the tiny baby.[43]

Importantly though, the background de Sitter space (or the regions of that space that are empty de Sitter) have no respective arrows of time. This is because empty de Sitter space is in a state of thermal equilibrium. Prior to universe nucleation, there is no entropic increase. Such a fact (noted by Carroll himself [2010], p. 355) makes interpreting Carroll's comments regarding the relationships between the arrows of time *per* metagalaxies, and the direction of time in the background space difficult to interpret, for he stated that "…local direction of time [*i.e.*, the direction of time in our metagalaxy] may not be related to that of the background space-time."[44] But again, with respect to the background space-time, or at least the appropriate regions thereof, there just is no direction of time. Something is awry.

    Is the FGG nucleation process governed by a time parameter? If it is, which time parameter is it? When we give attention to Carroll's writings, we see in them a clear commitment to the thesis that the nucleation process is in fact governed by a temporal metric or time parameter. For example, Carroll's illustration of the nucleation process in (Carroll ([2010], p. 357, Fig. 85) includes a time axis. That figure indicates that the process of FGG tunneling and metagalaxy nucleation occurs in time. In fact, Carroll believes that the background Universe increases its entropy through the nucleation of universes which themselves increase in thermal entropy, and this process of entropy increase is thought to be something which transpires *in time*. But which time? It cannot be a local time peculiar to the nucleated metagalaxy, for that entire space-time does not come into being until it pinches off near the end of the process. Likewise, the time parameter governing the Universe cannot be the time parameter governing the *entire* process of entropic increase *via* nucleation, since Carroll insists that on the heels of the pinching off stage of the process, one is left with two completely independent and autonomous space-times. Such independence is a consequence of the assumed mechanism of universe nucleation. FGG entails that no worldline can be drawn from mother to baby universe. In fact, for FGG-style mechanisms "no causal curve from the original phase can enter the new phase after the tunneling event…"[45] Thus, in order for the process to be one which occurs in time, a hyper or external time parameter is required.[46]

    The idea of an external time parameter is implausible. Carroll disapproves of the idea:

> The weirdest thing about the idea that the space of states changes with time is that it requires an *external* time parameter—a concept of "time" that lives outside the actual universe, and through which the universe evolves…There's not much to say about this idea. It's possible, but very few people advocate it as an approach to the arrow-of-time problem.²⁷⁴ It would require a dramatic rethinking of the way we currently understand the laws of physics; nothing about our current framework suggests the existence of a time parameter lurking outside the universe itself. So for now, we can't rule it out, but it doesn't give us a warm and fuzzy feeling.[47]

---

[43] (Carroll [2010], pp. 357-8 emphasis mine; *cf*. Carroll [2008b], p. 56).
[44] (Carroll [2006], p. 1134).
[45] (Aguirre, Gratton, and Johnson [2007], p. 123501-9). Their comments pertain to a generalization of the geometry of the FGG mechanism, what they call '"L" tunneling geometry'. Importantly, these authors go on to point out that "[h]olographic considerations would seem to conflict with the L geometries (at least for transitions to higher vacuum energy)…" [ibid.] Carroll takes the holographic principle seriously. He ([2010], p. 281) stated, "[t]he holographic principle is a very general idea; it should be a feature of whatever theory of quantum gravity eventually turns out to be right."
[46] The criticism is essentially Eric Winsberg's (*qq.v.* note 48 and 49 below).

A criticism akin to the one I have articulated here was voiced by Eric Winsberg.[48] Winsberg would no doubt agree, that if (as Carroll insists) the model entails a never ceasing increase in entropy (*in time*) through the nucleation of universes, then there is "an external time parameter, something Carroll explicitly, and correctly rejects…"[49]

A second inconsistency rears its head subsequent to reflecting upon the nature of the initial Cauchy hypersurface in the CC-M (and here I lean on Nikolić [2008], p. 2).[50] That initial hypersurface is thought to be generic. But this is not so. At every Cauchy hypersurface of the background space, save the *initial* Cauchy hypersurface, entropy increases away from that hypersurface out along a single direction in time. Only at the initial Cauchy hypersurface does entropy increase in two directions. And so I agree with Nikolić, "…the initial hypersurface having two directions of time is not typical at all."[51]

Although I will discuss scientific issues relevant to claim (2) below, I want to immediately point out a perceived ambiguity and incompleteness in Carroll's discussion of universe nucleation. First, I have already noted above, that Carroll (*qq.v.* p. 6, n. 35) interprets his work with Chen in such a way that it is committed to the quantum tunneling mechanism of FGG.[52] But something is amiss. In their original ([2004]) paper, C&C explicitly deny that their mechanism of nucleation involves any such quantum tunneling process. They stated:

---

[47] (Carroll [2010], pp. 341-2 emphasis mine).
[48] See (Winsberg [2012], pp. 401-2).
[49] (Winsberg [2012], p. 402).
[50] This second criticism applies to the CC-M, but not the ᴹCC-M, since if the entire background space is de Sitter, there are no arrows of time.
[51] (Nikolić [2008], p. 2). See also (Vilenkin [2013a], p. 21). Vilenkin ([2013b], pp. 20-21) admitted,
> "A generic spacelike hypersurface in this kind of spacetime will itself run into singularities, so an infinite regular Cauchy surface appears to be rather special. Note, by the way, that if one is willing to accept a spacetime besieged by singularities, then the assumption of an infinite Cauchy surface does not seem to be essential. A large compact Cauchy surface with generic initial data will also yield some inflating regions surrounded by singularities."

Vilenkin ([2013b], p. 21) also takes issue with C&C's insistence that the initial Cauchy hypersurface is generic. His worry is related to the question of whether or not, on the supposition that entropy is unbounded, any one state can truly be typical or generic. He said, "[i]f indeed the entropy of the universe is unbounded from above, then there is no such thing as a generic (or random, or typical) state." ([ibid.])

Putting worries about the initial time surface aside, I would like to add that it is difficult to make sense of the claim that any one unique set of states are generic since on C&C's view, everything that physically can happen, happens an infinite amount of times (see sect. 5.4 for more on this idea).

[52] Aguirre, (Sean) Carroll, and Johnson ([2011], pp. 22-3) provide a very detailed study of fluctuations and universe nucleation. They put pressure on recommended strategies for resolving likelihood worries akin to the Boltzmann Brain paradox which concern technical details about generating inflation "from a non-inflating phase" ([ibid.], p. 22). In the process of applying that pressure, they briefly describe one would-be escape. The authors intimate that the escape probably necessitates that observers cannot enter "the new inflating region….because the nucleated bubble is separated from the parent spacetime by a wormhole"; they state that "this is the Farhi-Guth-Guven process [74]" (Aguirre, Carroll, and Johnson [2011], p. 23). The authors go on to cite (Carroll and Chen [2004]), successfully associating that work with the quantum tunneling approach.

> In our discussion is that we [*sic*] examine the case of an *harmonic oscillator potential without any false vacua*; in such a potential we can simply fluctuate up *without any tunneling*. The resulting period of inflation can then end via conventional slow-roll, *which is more phenomenologically acceptable than tunneling from a false vacuum* (as in "old inflation" [7]). Thus, the emptying-out of the universe under typical evolution of a generic state can actually provide appropriate initial conditions for the onset of inflation, which then leads to regions that look like our universe.[53]

But C&C ([2004], pp. 22-23; pp. 25-26; *cf*. n. 4 on p. 26) concede that the fluctuation route to metagalaxy nucleation and large-scale structure formation is incredibly improbable.

I described the incompleteness of the model as "admitted incompleteness" because Carroll himself (with collaborators Aguirre and Johnson [2011], pp. 23-24) criticized the FGG mechanism for universe nucleation confessing (independently) in a different place that that mechanism is "extremely speculative".[54]

In other work, Carroll indicated that the multiverse is a prediction of string theory and inflation.[55] His optimism concerning string theory is somewhat surprising since "...there is presently no fully satisfactory embedding of de Sitter space into string theory"[56], and "[a]ll *explicit* and fully trustworthy solutions that have ever been constructed in string theory have a non-positive cosmological constant."[57] Andrew Strominger elaborated on this point:

> An obvious approach, successfuly [*sic*] employed in the black hole case, would be to begin by embedding de Sitter space as a solution of string theory, and then exploit various string dualities to obtain a microscopic description. Unfortunately persistent efforts by many (mostly unpublished!) have so far failed even to find a fully satisfactory de Sitter solution of string theory.[58]

---

[53] (Carroll and Chen [2004], p. 21 emphasis mine). FGG essentially involves false vacua. Moreover, the mechanism of (Garriga and Vilenkin [1998], also involve false vacua (Vilenkin [2013b], p. 12)).

[54] (Carroll [2006], p. 1133). It is important to remember how tentative the conclusions of Farhi, Guth, and Guven were. Recall their remarks at the end of the paper:
> "The inflationary universe model proposes that our universe grew from a tiny inflating region of false vacuum. We know, however, that the laws of classical general relativity imply that a bubble that grows large enough to become a new universe cannot be produced without an initial singularity. In this paper we have asked whether this requirement can be avoided by quantum tunneling. *Unfortunately we do not have a definitive answer to this question*, but we have obtained an expression for the tunneling amplitude that seems highly plausible, and we conjecture that it is a valid approximation." (Farhi, Guth, and Guven [1990], p. 472 emphasis mine; *cf*. p. 473)

[55] See (Carroll [2012]). See also his comments in (Carroll [2006], p. 1133), and the favorable attitude about string theory in his ([2010], pp. 284-6 "The leading candidate for a consistent quantum theory of gravity is *string theory*" ([ibid.], p. 284 emphasis in the original)). See Smolin ([2004]) for an evaluation of the merits of string theory over against loop quantum gravity (see particularly [2004], p. 521). I am a string theory skeptic who agreed with Tom Banks ([2007], p. 3) when he wrote that "[t]he Landscape…does not really give an explanation of how the universe gets into the low entropy state from which it tunnels into the basin of attraction in which we find ourselves."

[56] (Bousso, DeWolfe, and Myers [2003], p. 297-8). There are no-go theorems which seek to establish that certain compactified theories (string theories) are incompatible with de Sitter space-time (see Maldacena and Nuñez [2000], pp. 26-27).

[57] (Van Riet [2011], p. 2).

Captivatingly, Carroll (with Johnson, and Randall) seems to agree, "…string theory…seems to favor Minkowski or anti-de Sitter vacua."[59]

There are further problems with injecting string theory into the model, for that theory requires a great many dimensions which must somehow be compactified into any pure or asymptotically de Sitter space if one or the other is your space of choice. The problem is that there are no-go theorems proving that compactified theories which abide by the null energy condition (along with several other plausible conditions for string theoretic models) cannot be wed to inflationary theory.[60] It has also been shown that compactified theories which *violate* the null energy condition, but which otherwise satisfy other very plausible conditions (for string theoretic models) cannot be united with inflationary theory or theories.[61] So I'm not sure what to make of Carroll's claim that a multiverse is a prediction of inflation coupled with string theory. The two are not agreeable partners.[62]

The foregoing reasoning indicates that FGG nucleation out of a de Sitter space-time is merely speculative and that Carroll's discussion of it should be thought of as exploratory. I believe it is therefore safe to conclude that a central piece of the model is missing, and so the CC-M is incomplete in that it does not have a clear recommended dynamical path from the background Universe to the birth of metagalaxies like ours (*q.v.* note 41).

The incompleteness of the CC-M has a bearing on the question of whether or not the model provides a *bona fide* scientific explanation of our initial low-entropy state. Assuming some robust version of scientific realism, explanations, when they successfully explain, are at least approximately true. It is not clear how an *explanans* can be verisimilar, if it is unclear which proposition, if any, is expressed by that *explanans* on account of the kind of incompleteness the CC-M displays. Thus, I find this gap in the model to be severely delimiting. We cannot, in my opinion, justifiably claim that the CC-M proffers an actual scientific explanation of the initial non-empty smoothness of our metagalaxy, since it is altogether unclear what the explanation *is* on the CC-M.

---

[58] (Stominger [2001], p. 2). In fact, Stominger takes himself to be working in the context of a state of the art that is without "a stringy example of de Sitter space." ([ibid.])

[59] (Carroll, Johnson, and Randall [2009], p. 2). As the quotation from van Reit would seem to suggest, the supposition that string theory does not get along well with a pure de Sitter space-time may be related to the problem of the compatibility of string theory and space-times with a positive cosmological constant (the real presence of dark vacuum energy). On this issue Peebles and Ratra expressed an interesting thought,

> "Building on earlier work,[119] and Hellerman, Kaloper, and Susskind (2001) and Fischler *et al.* (2001) noted that dark-energy scalar field cosmological models have future event horizons characteristic of the de Sitter model. This means some events have causal futures that do not share any common events. In these dark-energy scalar field models, some correlations are therefore unmeasurable, which destroys the observational meaning of the *S* matrix. *This indicates that it is not straightforward to bring superstring/M theory into consistency with dark-energy models in which the expansion of the universe is accelerating.*" (Peebles and Ratra [2003], pp. 598-9 empahsis mine)

[60] I have in mind the results of Steinhardt and Wesley ([2009], pp. 104026-4 to 104026-6). Though *cf.* Koster and Postma ([2011]).

[61] Steinhardt and Wesley ([2009], pp. 104026-6 to 104026-8).

[62] See also Hertzberg et. al. ([2007]). They argue that inflationary theory will not run with the most intimately understood and perhaps most realistic compactifications of the string theoretic type IIA sort.

## 4   Unbounded Entropy?[63]

Having argued that C&C's proposed scientific explanation of our initial low-entropy state is incomplete, I now want to give attention to that explanation on the assumption that it is or can be completed. In this section, I take up claim (3). I maintain that the *N*-bound confirms the Tom Banks/Willy Fischler Λ-N correspondence thesis, at least when the background space of the $^M$CC-M is in view, and that such confirmation means that claim (3) is false. I also argue that while it is unclear if the *N*-bound holds for the background space of the CC-M, there are arguments to which one can turn for the purposes of establishing Λ-N correspondence for that space, and so claim (3) is false given the CC-M.

4.1 Λ-N Correspondence

Tom Banks has argued that the value of Λ, the cosmological constant, is the inverse of the value of *N*.[64] *N* is the logarithm of the dimension of Hilbert space in quantum theory. By consequence, if one's quantum theory conceives of *N* as finite, then that quantum theory will contain finitely many dimensions.[65] The correspondence of Λ to *N* entails that there is a large (though finite) number of degrees of freedom.[66] If, however, *N* really is finite, then quantum theories of gravity featuring infinitely many degrees of freedom will be implausible.

4.2 Confirmation of Λ-N Correspondence

Raphael Bousso has noted that proofs of what he calls the "*N*-bound" constitute evidence for Λ-*N* correspondence.[67] The *N*-bound states that every space-time with Λ > 0 is a space-time whose total *observable* entropy is bounded by:

(5): $N = \frac{3\pi}{\Lambda}$         (Eq. 2)[68]

Or, any space-time with a positive cosmological constant is one which cannot feature an observable entropy whose value is greater than $N = 3\pi/\Lambda$.[69] The *N*-bound trivially holds for empty de Sitter space-times like the background space of the $^M$CC-M. In addition, Bousso at one time believed that one could show that the *N*-bound is valid for asymptotically de Sitter space-times—such as our metagalaxy—on the basis of the generalized second law.[70]

---

[63] In this section, I use Planck units, and work with only four dimensions of space-time.

[64] See (Banks [2000], p. 5). He proffered three arguments for the view, though he was concerned with establishing the correspondence for asymptotically de Sitter space-times solely). For details regarding asymptotically de Sitter space-times see (Gibbons and Hawking [1977]).

[65] (Bousso [2000a], p. 2. n. 2).

[66] (Bousso [2000a], p. 2).

[67] His argument is explanatoral: "It is hard to see what, other than the Λ-N correspondence, would offer a compelling explanation [of] why such disparate elements appear to join seamlessly to imply a simple and general result" (Bousso [2000a], p. 18).

[68] (Bousso [2000a], p. 3). The type of entropy in play appears to be information-theoretic or Von Neumann entropy. This fact is irrelevant. The main argument of sect. 4.3 still runs.

[69] (Bousso [2000a], p. 2). In subsequent discussion, I will sometimes speak of *N*-bound validity for a space-time. What I mean by such a judgment is that Eq.2 (proposition 5) holds for those space-times.

Bousso attempted to establish the *N*-bound by connecting two further entropy bounds, *viz*., the *D*-bound, and covariant entropy bound.[71] The covariant entropy bound—developed for the purposes of helping along supporting argumentation for the holographic principle—is a bound on light-sheets or null hypersurfaces.[72] C&C believe that the covariant bound is implied by the holographic principle, and at least Carroll takes that principle seriously.[73] I will therefore omit an articulation of the supporting arguments for the covariant bound.

In order to understand the nature of the *D*-bound, several important notions in the literature on entropy require introduction. Many of these notions receive clarification and sound scientific study in the work of Jacob D. Bekenstein, who discovered that the total entropy of an asymptotically flat space is equivalent to the sum of the Bekenstein-Hawking entropy $S_h$, and the matter entropy $S_m$ of the space.[74] I should add (following Gibbons and Hawking [1977] and Bousso [2000a, p. 11-2]) that one should include the cosmological horizon entropy $S_c$—which, as it turns out, in an empty de Sitter space is just equivalent to *N* —in calculating the total entropy of an AsDS space-time. Now, the Bekenstein-Hawking entropy is black hole entropy. Cosmological horizon entropy is simply (and tautologically) the entropy of a cosmological horizon.[75] And matter entropy is the entropy of material bodies.

The notions appealed to for my statement of Bekenstein's discovery are just some of the prerequisite notions needed to understand the *D*-bound, for that bound also appeals to the generalized second law (GSL). The GSL states (roughly) that the total entropy of a system (which includes black-hole entropy) never decreases as time marches onward.[76] And so, with respect to an AsDS space, and the total cosmological horizon entropy $S_c$, and matter entropy $S_m$ of a system of that space, the GSL entails that the total cosmological horizon entropy of that

---

[70] Bousso ([2000a], p. 2) remarked:
> "It is not difficult to see that the *N*-bound is true for vacuum solutions like de Sitter space (a trivial case). Moreover, one can argue that it is satisfied for all space-times which are asymptotically de Sitter at late times, by the generalized second law of thermodynamics."

I provide a definition of the generalized second law in sect. 4.3 below.

As recently as 2012, Bousso ([2012], p. 29) confessed that "[t]he entropy bound in the corresponding de Sitter space is 3π/Λ...." This more current admission is important since it implies that his later change of mind regarding *N*-bound validity for all de Sitter space-times whether dS or AsDS, did not affect his belief that the *N*-bound holds for empty or pure dS space-times.

[71] (Bousso [2000a], p. 13). Bousso's proof of the *N*-bound also involves the notion of a causal diamond (for which see (Bousso [2000a], pp. 4-9)).

[72] (Bousso [2000a], p. 9; *cf*. Bousso [1999a], [2000a, pp. 10-1]; t' Hooft [2009]; and Susskind [1995]). C&C show a certain respect for the covariant bound, in that they finesse their model so as not to violate it (see the comments in (Carroll and Chen [2004], pp. 14-5)). Banks ([2007], p. 19) notes that you can prove the covariant entropy bound "from Einstein's equations with additional assumptions bounding entropy density by energy density"

[73] "A concrete consequence of the holographic principle is Bousso's covariant entropy bound, which places a limit on the entropy that can be contained within a region [63]." (Carroll and Chen [2004], p. 14) With respect to Carroll's attitude concerning the holographic principle, see (Carroll [2010], pp. 278-81). For more on the holographic principle, see (Bousso [2000b]). For discussion of holographic cosmology, see (Banks [2010], pp. 4875-7; Banks and Fischler [2001]; and Fischler and Susskind [1998]).

[74] (Bousso [2000a], pp. 11-2; and for background see Bekenstein [1972], [1973], [1974] following Bousso's source trail).

[75] What's a cosmological horizon? It's the horizon of a postulated observer. Cosmological horizons are sometimes called particle horizons.

[76] See (Wall [2009], p. 2).

system will be greater than or equal to the matter entropy of the system (where black hole entropy is accounted for in the matter entropy).[77] Thus, $\Delta S_c \geq S_m$.[78] The *D*-bound then, is a bound on matter entropy which marks the "[d]ifference between *N* and the horizon entropy" of a system.[79]

Given the above details, Bousso's ([2000a]) proof of the *D*-bound is as follows: Suppose there is a system of matter σ situated in an asymptotically de Sitter space-time that is enveloped within the area of a cosmological horizon $A_c$. Suppose further that σ is headed—evolution-wise—to an empty de Sitter state. If we posit the existence of a hypothetical observer within our assumed matter system, the evolution of σ toward an empty de Sitter space-time can be illustrated by simply noting that the observer will be moved, as the evolution of the system marches forward, into the asymptotic region (the de Sitter region). Eventually our hypothetical observer will start to think that σ is moving into the cosmological horizon, and as σ evolves in this way, $S_m$ will vanish, though the cosmological horizon entropy will be enlarged by the following quantity[80]:

$$(6): \Delta S_c = \frac{1}{4}(A_0 - A_c) \qquad \text{(Eq. 3)}^{81}$$

But now—Bousso insists—it will follow, given $A_0 = 4N$, that there is a bound on $S_m$, *viz.*, the *D*-bound:

$$(D\text{-Bound}): S_m \leq N - \frac{1}{4}A_c \qquad \text{(Eq. 4)}^{82}$$

With the *D*-bound and covariant bound in his back pocket, Bousso only needed a little more equipment (*i.e.*, two fairly non-contentious results regarding causal diamonds[83]) to prove that the *N*-bound is valid for all space-times with a positive cosmological constant.[84]

The tools in Bousso's back pocket are only needed for generalizing Bousso's proof of the *N*-bound to all space-times with a positive cosmological constant.[85] Lee Smolin ([2002, p. 45]) articulated a straightforward proof of the validity of the *N*-bound for empty de Sitter space-times, which required only the Bekenstein bound.[86] Moreover, he derives the *N*-bound from implications of loop quantum gravity in (Smolin [2002], pp. 45-46). This latter result is evidence that the *N*-bound is valid for empty de Sitter space in light of a possible, though in some ways still incomplete, fundamental theory of gravity.[87]

---

[77] Following (Bousso [2000a], p. 12).
[78] Following (Bousso [2000a], pp. 12-3).
[79] (Bousso [2000a], p. 13). This bound is obviously only relevant when non-empty de Sitter space-times are in view.
[80] Again, the argument here is from (Bousso [2000a], p. 12)
[81] (Bousso [2000a], p. 12).
[82] (Bousso [2000a], p. 12). Bousso in [ibid.], p. 18, and in [2001] noted that the *D*-bound is akin to the Bekenstein bound for which see Bekenstein ([1981]).
[83] See (Bousso [2000a], pp. 5-9).
[84] For the proof see (Bousso [2000a], pp. 13-7).
[85] I am unsure of whether or not the *N*-bound is valid for all space-times with a positive cosmological constant. Indeed, Bousso himself (with collaborators) provides counter-examples to the *N*-bound (see Bousso, DeWolfe, and Myers [2003]). These counter-examples involve space-times with dimensionality greater than four. Clarkson, Ghezelbash, and Mann ([2003]) attempted to show that the *N*-bound is not valid for a four-dimensional Taub-Bold space-time. The Taub-Bold space-time they had in mind is locally asymptotically de Sitter, and it features NUT charge (magnetic mass), and (unfortunately) closed timelike curves ([ibid.], pp. 360-1). This does not appear to be the background space-time of the $^M$CC-M or CC-M. With respect to *N*-bound validity, the only point that my argumentation requires is that the *N*-bound is valid for dS or empty de Sitter space-time, and both Bousso ([2000a], [2012]) and Smolin ([2002]) have acknowledged its validity in that context.

4.3 The *N*-Bound and the $^M$CC-M

How does all of this relate to the $^M$CC-M? Recall proposition (3) above, and remember that if (3) holds, then there are infinitely many degrees of freedom.[88] The *N*-bound, which is trivially valid for empty de Sitter space (the very background space of $^M$CC-M) is strong confirming evidence for the Banks/Fischler $\Lambda$-*N* correspondence thesis (as Bousso suggested). But educe from your memory the fact that *N* comports to the logarithm of the dimension of the Hilbert space in quantum theory. If the correspondence thesis is right, then *N* is probably finite. Therefore, there should be finitely many dimensions of Hilbert space in the correct quantum theory, and so there are also only finitely many degrees of freedom. This conclusion ensures that (3) is false. Entropy is not unbounded from above.[89] The argument in play can be summarized as follows:

(Premise 1): If the $\Lambda$/*N* correspondence thesis holds for dS space-time, then the correct quantum theory describing that space-time will feature a finitely dimensional Hilbert space.
(Premise 2): If the *N*-bound is valid for dS space-time, and the best explanation of *N*-bound validity for dS space-time is the $\Lambda$/*N* correspondence thesis, then the $\Lambda$/*N* correspondence thesis holds for dS space-time.
(Premise 3): The *N*-bound is valid for dS space-time, and the best explanation of *N*-bound validity for dS space-time is the $\Lambda$/*N* correspondence thesis.

---

[86] See (Pesci [2010]) for an argument from holography for the Bekenstein bound. In addition, Smolin ([2002], p. 45) states that the *N*-bound is valid "[f]or a semiclassical quantum field theory in deSitter spacetime" given only "Bousso's form of the holographic bound".

[87] I do not mean to suggest that loop quantum gravity is the correct complete theory of quantum gravity. Nor do I mean to suggest that loop quantum cosmology provides approximately true models of the cosmos. My point is simply that loop quantum gravity is at least a potential approximation of what a complete quantum gravity will look like, and since it implies that the *N*-bound is valid for empty de Sitter space, we have some evidence that such validity will hold in light of quantum gravity.

[88] C&C's commitment to the thesis that there are infinitely many degrees of freedom and that this thesis is connected to (3) is clear. They stated:
> "…there is one loophole in this reasoning, namely the assumption that there is such a thing as a state of maximal entropy. If the universe truly has an infinite number of degrees of freedom, and can evolve in a direction of increasing entropy from any specified state, then an explanation for the observed arrow of time arises more naturally." (Carroll and Chen [2004], p. 7; *cf*. pp. 14-5, and p. 30. In fact on page 15 they state that it is an assumption of the portion of their paper where the model is articulated that there are an infinite amount of degrees of freedom.)

[89] Notice that the argument is not the claim that (3) is false because there is a bound which bounds the entropy of C&C's background space. I should add that in (Aguirre, Carroll, and Johnson [2011], p. 10) it is admitted that "the fundamental degrees of freedom underlying dS space are unknown." They also claim that "[c]omplementarity plus the bound on the information accessible to any one observer…implies that dS can be described by a theory with a finite number of degrees of freedom…" ([ibid.])

(Premise 4): If the correct quantum theory for an empty dS space-time features a finitely dimensional Hilbert space, then dS space-time features only finitely many degrees of freedom.
(Premise 5): If dS space-time features only finitely many degrees of freedom, then the global entropy of dS space-time cannot be unbounded from above.
(Conclusion): Therefore, the global entropy of dS space-time cannot be unbounded from above.

The first premise is true by virtue of the meaning of the correspondence thesis. The second premise holds on account of the cogency of inference to the best explanation reasoning. In the absence of defeaters and underdetermination, such reasoning provides cognizers with epistemic justification for their belief that the purported best explanation holds.[90] The first conjunct of premise three follows from the insights and considerations of sect. 4.2. The second conjunct follows from the fact that there is simply no competing explanation of the relating of the two seemingly incommensurable parameters, *viz.* $\Lambda$ and $N$ (*q.v.*, note 67). It seems that the correspondence thesis wins by default. Premises four and five seem straightforward enough, and our conclusion follows from elementary moves in propositional logic.

In an attempt to defend the $^M$CC-M, one might respond by emphasizing the fact that the means by which the Universe increases its entropy is by giving birth to metagalaxies (*q.v.*, note 39).[91] Appeals to the *N*-bound do nothing to subvert that possibility. This response is flawed. According to C&C, if it is not the case that there are infinitely many degrees of freedom, then their story regarding universe nucleation and unbounded entropy cannot run. Entropy is unbounded from above *only if* there are infinitely many degrees of freedom. The above argumentation cuts down this necessary condition, and so results in a bound on entropy.

Again the argument from the *N*-bound shows that with respect to the background de Sitter space-time of the $^M$CC-M, there are finitely many degrees of freedom. Carroll himself believes that the $^M$CC-M would in that case have a fundamental problem with Poincaré recurrence.[92] Recall that on the basis of Newtonian mechanical laws of motion, and with respect to an energetically isolated system whose volume is finite, Poincaré proved an important theorem. The result is this: given the aforementioned assumptions, a relevant system which starts off in state *s* at *t*, will, given enough time, evolve back into a state arbitrarily close to *s*, and it will do this infinitely many times.[93] There are quantum analogs of this theorem,[94] and Carroll believes he can escape these analogs by appeal to an infinitely dimensional Hilbert space.[95] But

---

[90] The use of this type of explanatory reasoning for the purposes of establishing the Banks/Fischler correspondence thesis was used by Bousso (*q.v.*, note 67).

[91] (Carroll [2010], pp. 359-360).

[92] See (Carroll [2008b], pp. 6-7)

[93] I'm relying upon and paraphrasing the discussion in (Sklar [1993], p. 36). Poincaré stated, "[a]ny phase-space configuration (*q, p*) of a system enclosed in a finite volume will be repeated as accurately as one wishes after a finite (be it possibly very long) interval of time." as quoted by (Bocchieri and Loinger [1957], p. 337).

[94] Dyson, Kleban, and Susskind ([2002], p. 17) put the quantum version this way:
> "The quantum Poincaré Recurrence theorem…can be stated as follows: given a system in which all energy eigenvalues are discrete, a state will return arbitrarily close to its initial value in a finite amount of time. It follows immediately that expectation values of observables will also return arbitrarily close to their original values."

See also Bocchieri and Loinger ([1957]); Duvenhage ([2002], pp. 53-60); Schulman ([1978]); *cf.* the discussion in Percival ([1961]).

[95] See (Carroll [2008b], pp. 6-7).

you will remember that the argument from the *N*-bound cuts down the dimensions of Hilbert space to only a finite amount due to the Banks/Fischler $\Lambda$-*N* correspondence thesis. Thus, by Carroll's own lights, the problem of Poincaré recurrence remains.

Let me now grant, for the sake of argument, that the entropy increase through universe nucleation-response muzzles the argument from the *N*-bound; there still remains an insurmountable problem. Carroll understands metagalaxies to be autonomous independent space-times. Respective proto-inflationary patches (those patches that produce the large-scale structure of metagalaxies like ours) do relate in some—heretofore unknown—causal way to prior fluctuations or tunneling in/from the background space. In addition, the metagalaxy evolves unitarily as an isolated causally disconnected metagalaxy. I am afraid that I do not understand what it means to say that the entropy of the Universe increases "in time" (which time we do not know) as the entropy of the causally independent and autonomous space-time that is a metagalaxy increases. My worry here is not about the problem of time and nucleation (discussed in sect. 3), but about the following peculiar relationship: as entropy increases in an independent and autonomous metagalaxy *m*, entropy increases in the background mother Universe. I do not believe, therefore, that what Carroll has "done is given the [U]niverse a way that it can increase its entropy without limit."[96]

4.4 The *N*-Bound and the CC-M

Does the argument from the *N*-bound apply equally well to the background space-time of the CC-M? I am not sure. C&C's description of that space is fragmented. We do not know the dimensionality of the space, nor what generic conditions the space evolves away from. In addition, we do not know what precise kinds of matter occupy the space in its non-de Sitter regions. Ignorance of these matters makes it difficult to determine *N*-bound validity, for although Bousso ([2000a]) originally argued that the *N*-bound is valid for all space-times with a positive cosmological constant, he would later (with collaborators) reverse his opinion on the matter by proffering counter-examples to his original proof. These counter-examples all come from space-times with dimensionality greater than four, and from space-times which violate a particular "assumption of spherical symmetry."[97] But let us suppose that the *N*-bound is not valid for the background space of the CC-M. Tom Banks ([2000], pp. 5-6) provided three convincing arguments all demonstrating that the $\Lambda$-*N* correspondence thesis holds for AsDS space-times. From the little we can discern about the nature of the background space of the CC-M, we can somewhat safely infer that that space is AsDS. Hence, the Hilbert space of the appropriate quantum theory describing that space-time is finitely dimensional. Claim (3) is therefore false when either the CC-M or $^M$CC-M is in view.

I will now continue to assume that the CC-M/$^M$CC-M[98] is complete, and move on and reflect, in the next section, on claim (2), evaluating the proposed mechanisms for universe nucleation in the work of C&C.

---

[96] (Carroll [2010], p. 259).
[97] (Bousso, DeWolfe, and Myers [2003], p. 299). *Q.v.* note 85
[98] Throughout the remainder of the paper, one may read '$^M$CC-M' wherever one sees 'CC-M'. All subsequent argumentation will be applicable to both.

## 5 Nucleation and Metagalaxy Creation

As I have already pointed out, Carroll seems to commit himself to the quantum tunneling process of universe nucleation as articulated by FGG ([1990]).[99] That process will not serve as a proper mechanism for the nucleation of our metagalaxy, if our metagalaxy has an *initial* singularity. On this point FGG stated, "…any plausible scheme to create a universe in the laboratory must avoid an initial singularity."[100] As a result, FGG try to articulate a theory of quantum tunneling which avoids the Penrose singularity theorem of ([1965]). I will argue that while the FGG model may escape the Penrose theorem, it does not escape other theorems which entail that our metagalaxy has an initial singularity, and that our metagalaxy is past-geodesically incomplete.

5.1 The EGS Theorem and Related Results

According to the EGS theorem (proven in Ehlers, Geren, and Sachs ([1968])), given the Copernican principle[101], and that observers situated in some expanding model discern (*via* observations) that free and unrestrained "propagating background radiation is" isotropic, the space-time in which such observers are situated must be FLRW.[102] Clifton, Clarkson, and Bull ([2012]) (CCB) showed that space-time geometry is, for an observer, FLRW "using the CMB alone" without the Copernican principle.[103] Their proof also indicates that "our *entire* causal past must…be FLRW."[104] One acquires their results by assuming that an observer beholds isotropic cosmic microwave background radiation while the Sunyaev-Zel'dovich effect ((SZ) which involves baryonic matter scattering the photons of the CMBR[105]) is present in that beholding.[106] The idea is that if a single onlooker observes blackbody CMBR that is isotropic, and that CMBR is accompanied by particular SZ-related scattering events, then that observer can infer that her universe is FLRW, given that the necessary assumptions of the EGS theorem (save the Copernican principle) hold, and that either (a) the observer's observations are over a prolonged period of time, or (b) the SZ-related effects involve double scattering.[107] I should add that the CCB results hold even given the presence of dark energy, it is just that such dark energy must be susceptible to a scalar field description.[108]

---

[99] Again, see (Carroll [2008b], p. 8, [2010], pp. 356-9, and Aguirre, Carroll, and Johnson [2011], pp. 22-3).

[100] (Farhi, Guth, and Guven [1990], p. 419). Farhi and Guth ([1987], p. 149 stated, "[t]he requirement of an initial singularity appears to be an insurmountable obstacle to the creation of an inflationary universe in the laboratory.")

[101] The Copernican principle says, roughly, that our causal past and position in space-time is not unique or distinctive. (Stoeger, Maartens, and Ellis [1995], p. 1).

[102] Borrowing some wording from Smeenck ([2013], pp. 630-1). See also (Stoeger, Maartens, and Ellis [1995], p. 1). There is a nice discussion of these matters in (Clarkson and Maartens [2010]; Maartens [2011]; and Weinberg [1972], pp. 395-403, *cf.* [2008], p. 3). It is important to add that the result from Ehlers, Geren, and Sachs does not extend to times prior to the decoupling era. ([1968], p. 1349 "the result presented cannot be taken to mean that the universe *in its earliest stages* was necessarily a Friedmann model…" emphasis mine)

[103] (Clifton, Clarkson, and Bull ([2012], p. 051303-4).

[104] (Clifton, Clarkson, and Bull ([2012], p. 051303-3) emphasis mine.

[105] (Clarkson [2012], p. 19).

[106] (Clifton, Clarkson, and Bull [2012], pp. 051303-1 to 051303-2). For more on the Sunyaev-Zel'dovich effect see (Weinberg [2008], pp. 132-5).

[107] I'm paraphrasing Clarkson's review of the CCB result in (Clarkson [2012], p. 19).

Both the EGS and CCB results are significant since our observations regarding the cosmic microwave background radiation suggest that that blackbody radiation is *nearly* isotropic.[109] The qualifier 'nearly' is important since it seems that both EGS and CCB reasoning require highly idealized propagating radiation in so far as that radiation must be exactly isotropic.[110] Our metagalaxy's CMBR exhibits certain anisotropies[111], and so it is unclear what work these theorems can do for me.[112]

There are related results which do not rely on a condition of perfectly isotropic CMBR. One attempt to stabilize the EGS theorem in light of the inexact isotropy of the CMBR comes to us from the work of Stoeger, Maartens, and Ellis ([1995]).[113] They argued that our cosmos is approximately or nearly FLRW given the Copernican principle, the fact that background blackbody radiation is freely propagating everywhere and that such radiation is perceived, by observers, to be approximately or nearly isotropic (plus a few additional technical assumptions).[114] Maartens and Matravers ([1994]) have articulated a matter analog of the EGS theorem. Their result establishes that our universe is FLRW given the Copernican principle, and that a class of galactic observations along a postulated observer's world line is isotropic.[115]

The most formidable EGS-like result was recently discussed by Roy Maartens ([2011], pp. 5121-5) in his excellent review of much of the associated literature.[116] The theorem has it that with respect to a region of a space-time featuring dark energy (whether understood in terms of a

---

[108] It may be that in order to alleviate worries about fine-tuning and the cosmological constant, one should appropriate a scalar field model of dark energy. In addition, it seems that the best way of understanding dark energy *via* quintessence is to posit a scalar field model of dark energy. As Weinberg remarked, "[t]he natural way to introduce a varying vacuum energy is to assume the existence of one or more scalar fields, on which the vacuum energy depends, and whose cosmic expectation values change with time." (Weinberg [2008], p. 89) For more on dark energy and scalar field models of such energy, see (Sahni [2002], pp. 3439-41).

[109] Clarkson and Maartens ([2010], p. 2) stated,
> "Isotropy is directly observable in principle, and indeed we have excellent data to show that the CMB is isotropic about us to within one part in $\sim 10^5$ (once the dipole is interpreted as due to our motion relative to the cosmic frame, and removed by a boost)."

Weinberg ([2008], p. 129) confesses that treating the CMBR as perfectly isotropic and homogeneous is "a good approximation". He says that "the one thing that enabled Penzias and Wilson to distinguish the background radiation from radiation emitted by earth's atmosphere was that the microwave background did not seem to vary with direction in the sky." ([ibid.])

[110] Clifton, Clarkson, and Bull admit to their idealized assumptions in (Clifton, Clarkson, and Bull [2012], p. 051303-4]).

[111] See (Hawking and Ellis [1973], pp. 353-4). For a discussion of the CMBR anisotropies, see (Lyth and Liddle [2009], pp. 152-69; and Weinberg [2008], pp. 129-48).

[112] Ehlers, Geren, and Sachs ([1968]) also ignored the cosmological constant.

[113] See also the discussion in (Peebles [1993]).

[114] Technically the result is that:
> "*if* the Einstein-Liouville equations are satisfied in an expanding universe, where there is present pressure-free matter with 4-velocity vector field $u^a$ ($u_a u^a = -1$) such that (freely propagating) background radiation is everywhere almost-isotropic relative to $u^a$, *then* spacetime is almost FLRW." (Stoeger, Maartens, and Ellis [1995], p. 1 emphasis in the original) perfect fluid, quintessence, or cosmological constant) and dust matter, *if* (a) the Copernican principle holds, (b) the observed CMBR "rest frame is geodesic"[117] with an expanding four-velocity, and (c) the self-same radiation is collisionless with a vanishing octupole, quadrupole and dipole[118], *then* the *metric* of the relevant spacetime is FLRW.[119] The assumptions of this theorem are quite weak. I therefore agree with Maartens "[t]his is the most powerful observational basis that we have for background homogeneity and thus an FLRW background model."[120]

[115] These galactic observations correspond to propositions (O1)-(O4) in (Maartens and Matravers [1994], p. 2694). They are *not* observations of isotropic background blackbody radiation. See also (Maartens [2011]; and *cf.* Hasse and Perlick [1999]).

[116] His discussion of the specific result I am interested in is an expansion on his earlier work with Chris Clarkson in (Clarkson and Maartens [2010]).

What is the relevance of all of this to the CC-M? It turns out that every FLRW model (with matter like ours) features an *initial* singularity.[121] And since the assumptions of several of the EGS-like results are quite weak, we are justified in maintaining that our metagalaxy is best described by an FLRW model.[122] Thus, the FGG mechanism for metagalaxy nucleation cannot be the mechanism responsible for our universe's nucleation out of a background de Sitter space. Some other theory of nucleation that is not impeded by the singular nature of our metagalaxy is required.

5.2 The BGV Theorem

On the standard hot big bang model, implications of proper solutions to Einstein's field equations imply that our metagalaxy is geodesically incomplete in that our metagalaxy features an initial singularity. Attempts to avoid this implication were blocked by work on singularity theorems in the 1960s and 1970s. For example, Robert Geroch ([1966]), Stephen Hawking ([1965], [1966a], [1966b], [1967]) and Roger Penrose ([1965]) showed that any time-oriented space-time which satisfies modest conditions will be time-like or null geodesically incomplete.[123] In ([1970]) Hawking and Penrose attempted to generalize on this work by advancing "[a] new theorem on space-time singularities".[124] Hawking would later describe this newer theorem as one which predicts that there are singularities in the future, and that there is a singularity in the past "at the beginning of the present expansion of the universe."[125] The theorem had need of four

---

[117] (Maartens [2011], p. 5131]).

[118] Such that $F_\mu = F_{\mu\nu} = F_{\mu\nu\alpha} = 0$ holds (from equation 3.24 of Maartens [2011], p. 5125).

[119] See (Maartens [2011], p. 5125, p. 5131).

[120] (Maartens [2011], p. 5125).

[121] "FLRW models with ordinary matter have a singularity at a finite time in the past." (Smeenk [2013], p. 612). Hawking and Ellis stated, "…there are singularities in any Roberston-Walker space-time in which $\mu > 0$, $p \geq 0$ and $\Lambda$ is not too large." (Hawking and Ellis [1973], p. 142). See also (Wald [1984], pp. 213-4); and the discussion of FLRW models in (Penrose [2005], pp. 717-23). Subsequent to illustrating the class of FLRW models *via* Fig. 27.13, Penrose wrote,

> "Friedmann-Lemaitre-Roberston-Walker (FLRW) [are] spatially homogenous and isotropic cosmological models. …*each model starts with a Big Bang* [emphasis mine here]…In Figure.27.13a,b,c, I have tried to depict the time-evolution of the universe, according to Friedmann's original analysis of the Einstein equation, for the different alternative choices of spatial curvature. In each case, the universe starts form a *singularity* [emphasis in the original]—the so-called Big Bang—where spacetime curvatures become infinite and then it expands rapidly outwards." (Penrose [2005], p. 719) seemingly modest conditions, one of which demanded that space-time be described by Einstein's field equations along with a cosmological constant that is non-positive (*i.e.*, negative or equal to zero in value). It turned out that this modest condition was not modest enough. When inflationary stages of cosmic evolution are added to the standard model, a positive cosmological constant is required, thus, the Hawking-Penrose theorem "cannot be directly applied" to such models.[126]

[122] Stephen Hawking and G.F.R. Ellis ([1973], pp. 358-9) have argued similarly.

[123] See the review of many of these theorems in (Hawking and Ellis [1973], pp. 261-75).

[124] (Hawking and Penrose [1970], p. 529). This paper also provides an excellent review of both Hawking and Penrose's previous work on singularity theorems (see especially [ibid.], pp. 529-33).

[125] (Hawking [1996], p. 19).

Later theorems were proven. One of these was a result of the work of Arvind Borde and Alexander Vilenkin ([1996], pp. 819-22). They showed that a space-time is past-null geodesically incomplete if that space-time satisfies what were perceived to be even more modest conditions than those used to deliver erstwhile singularity theorems.[127] One such condition (*viz.*, the null convergence condition which is implied by the weak energy condition) was shown to be problematic in light of diffusion regions, and so that condition was not mild enough.[128]

Borde and Vilenkin would later return, this time with Alan Guth, to prove a newer theorem.[129] The Borde-Guth-Vilenkin (BGV) theorem entails that all space-times whose Hubble parameters are on average greater than zero, are past-geodesically incomplete.[130] Notice that the theorem does not necessarily suggest that the relevant space-times feature an initial singularity. This is because the theorem is not actually a *singularity* theorem.[131] The theorem only implies that every past-null or past-timelike geodesic is such that it cannot extend further than a past-boundary $\mathcal{B}$.[132]

The BGV has broad application potential since it only relies on a single, model independent assumption. For example, Borde, Guth, and Vilenkin apply the theorem to the early cyclic cosmogenic model of Steinhardt and Turok ([2002a]).[133] They also apply the theorem to a particular *part* of the ultra-large-scale structure in the higher-dimensional model of Martin Bucher ([2002]). This latter application is *apropos* because it is very loosely analogous to an application of the BGV to our independent nucleated metagalaxy on the CC-M.[134] One need not apply the BGV to ultra-large scale structure *in toto*.

---

[126] The quoted bit is from (Hawking and Penrose [1970], p. 531). Of course, they were not concerned with inflationary cosmology in 1970. Here is the broader context of the quote, "…we shall require the slightly stronger energy condition given in (3.4), than that used in I. This means that our theorem cannot be directly applied when a *positive cosmological constant* λ is present." (Hawking and Penrose [1970], p. 531 emphasis in the original). Many authors have noted that inflationary cosmological models violate the strong energy condition (the condition having to do with the value of λ) of the Hawking-Penrose theorem. See, for example, (Wall [2013], pp. 25-6. n. 13; and Borde and Vilenkin [1996], p. 824. n. 17), *inter alios*.

[127] The three conditions are stated in (Borde and Vilenkin [1996, p. 819]).

[128] In fact, Borde and Vilenkin themselves admitted that the weak energy condition is violated in certain space-time regions, when in those regions quantum fluctuations of the inflaton field dominate respective dynamics. These are the diffusion regions of the relevant space-times (Borde and Vilenkin [1997], p. 718). The weak energy condition implies the null convergence condition, and so if the latter condition does not hold in some space-time region, neither does the former. Borde and Vilenkin would also determine that an averaged or integral form of the null convergence condition is of no use in bypassing the diffusion problem (Borde and Vilenkin [1997], p. 719-20).

[129] (Borde, Guth, and Vilenkin [2003]).

[130] "The result depends on just one assumption: The Hubble parameter $H$ has a positive value when averaged over the affine parameter of a past-directed null or noncomoving timelike geodesic." (Borde, Guth, and Vilenkin [2003], p. 151301-4). See also (Mithani and Vilenkin [2012], p. 1 "…it [the BGV] states simply that past geodesics are incomplete provided that the expansion rate averaged along the geodesic is positive: $H_{av} > 0$."); and (Vilenkin [2013a], [2013b], p. 2]).

[131] See (Agullo, Ashtekar, and Nelson [2013], p. 2; Easson, Sawicki, and Vikman [2013], p. 1; and Guendelman and Steiner [2013], p. 1) who all suggest that the BGV demonstrates that inflationary space-times have initial singularities. This is not right. The point I'm making here was made by Vilenkin in his ([2013a], [2013b], p. 2).

[132] (Vilenkin [2013a], p. 2 "All it [the BGV theorem] says is that an expanding region of spacetime cannot be extended to the past beyond some boundary $\mathcal{B}$. All past-directed timelike and null geodesics, except perhaps a set of measure zero, reach this boundary in a finite proper time (finite affine parameter in the null case).")

[133] (Borde, Guth, and Vilenkin [2003], p. 151301-4); Guth [2004], p. 49 stated,

Our space-time or metagalaxy is such that it can be accurately described with a Hubble constant whose value is on average greater than zero. Hence, the BGV theorem can be easily applied to our metagalaxy. This point is underscored by the fact that the BGV was originally developed for the purposes of demonstrating that inflationary models are past-incomplete. Carroll and Chen are fans of inflation (*a fortiori* eternal inflation). They believe that in the past our metagalaxy exhibited an extraordinary inflationary stage of cosmic evolution. And so the theorem should be easily applicable to our metagalaxy as understood by the CC-M.

Is the presence of a past-boundary indicative of an initial singularity? For my present intents and purposes, it is. FGG define an initial singularity as "…a point on the boundary of space-time at which at least one backward-going (maximally extended) null geodesic terminates."[135] The BGV entails such geodesic incompleteness given that our metagalaxy satisfies the Hubble parameter condition (which on the CC-M it does).

C&C discuss the BGV theorem, citing (Borde, Guth, and Vilenkin [2003]) and interpreting the result in such a way that it suggests that eternal inflationary models have singulari*ties*.[136] This reading of the theorem is multiply flawed (*q.v.*, note 131).[137] C&C seem to imagine that because singularities "occur all the time at the center of black holes, and eventually disappear as the black hole evaporates" the BGV is unproblematic for their model.[138] They go on to remark that the fact that the theorem entails the presence of singularities does not itself entail that there is a "spacelike" boundary "for the entire spacetime."[139] Again, this is just a misstatement of the result. The theorem implies the existence of just such a boundary (as Vilenkin himself noted).[140] An interesting, *separate* question is whether or not the BGV applies to the Universe, or to our metagalaxy given the CC-M. I have argued that it at least applies to our metagalaxy.

---

"One particular application of the theory [he has in mind the BGV theorem] is the cyclic ekpyrotic model of Steinhardt & Turok ([2002]). This model has $H_{av} > 0$ for null geodesics for a single cycle, and since every cycle is identical, $H_{av} > 0$ when averaged over all cycles. The cyclic model is therefore past-incomplete and requires a boundary condition in the past."

See also (Mithani and Vilenkin [2012], pp. 1-2).

[134] Keep in mind that on the CC-M, our metagalaxy is an autonomous, independent space-time. Inquiring about whether or not the BGV applies to our metagalaxy and not the entire Universe makes sense.

[135] (Farhi, Guth, and Guven [1990], p. 419).

[136] (Carroll and Chen [2004], p. 27. n. 6).

[137] Vilenkin stated, ([2013b], p. 2) "[e]ven though the BGV theorem is sometimes called a 'singularity theorem', it does not imply the existence of spacetime singularities."

[138] (Carroll and Chen [2004], p. 27. n. 6).

[139] Ibid.

[140] Susskind interprets the results of the BGV theorem accurately. He ([2012a], p. 3, *cf*. [2012b]) provided a "for all intents and purposes"-response to the theorem, noting that "…in any kind of inflating cosmology the odds strongly (infinitely) favor the beginning to be so far in the past that it is effectively at minus infinity." (The latter part of this sentence is a quotation of Susskind [2012a], p. 3, not the former). While I believe that in the context of the CC-M, Susskind's worry would be relegated to ultra-large scale structure, *i.e.*, the question of whether or not the Universe is past-geodesically incomplete, I believe that a proper response to him need not go beyond, "so what?".

5.3 Evasion by the Quantum?

What about escaping the singularity and geodesic incompleteness *via* the quantum? Surely there is some hope that a more complete cosmogenic model outfitted with a full-blooded quantum understanding of gravity will consign our metagalaxy's initial singularity and past boundary to the trash bin of physical cosmology. McInnes reports that "[i]t has been argued…that quantum-mechanical effects allow the singularity in the Farhi-Guth 'wormhole' to be evaded..."[141] Carroll has expressed similar optimism.[142] Sadly however, quantum cosmogony does not justify such optimism.[143] There is no piece of classical cosmology on which the BGV theorem essentially relies, and for which we have *sufficient* evidence that that piece will be completely done away with in the quantum regime. In other words, the BGV theorem does not assume a classical theory of gravity. Vilenkin made this point clear:

> A remarkable thing about this theorem is its sweeping generality. We made no assumptions about the material content of the universe. We did not even assume that gravity is described by Einstein's equations. So, if Einstein's gravity requires some modification, our conclusion will still hold. The only assumption that we made was that the expansion rate of the universe never gets below some nonzero value, no matter how small. *This assumption should certainly be satisfied in the inflating false vacuum*. The conclusion is that past-eternal inflation without a beginning is impossible.[144]

But what about my use of results which capitalize on the EGS theorem and related reasoning? Are not those results classical? Yes, the results are classical. They depend upon the assumption that Einstein's field equations describe the cosmos.[145] However, we have no conclusive evidence that these results will be overturned by a complete quantum cosmology.

Perhaps you are still dissatisfied with my argumentation. The question, "how can we be sure that there is an initial space-time singularity at $\mathcal{B}$ in a full quantum physical context?" may still strike you as a deep worry.[146] I believe I can mollify the force of such a worry, since Aron Wall ([2013]) has recently proven a *quantum* singularity theorem that relies only upon the generalized second law (GSL),[147] which states that *generalized* entropy never decreases as time marches forward.[148]

---

[141] (McInnes [2007], p. 20, who cites Fischler, Morgan, and Polchinski [1990]; though *cf*. Vachaspati [2007]).

[142] See (Carroll [2008a], p. 4, [2010], p. 50, pp. 349-50, particularly p. 408. n. 277 "Also, the concept of a 'singularity' from classical general relativity is unlikely to survive intact in a theory of quantum gravity."), but *cf*. (Penrose [1996], p. 36) for a different view.

[143] In his final analysis, Carroll ([2010], p. 350) admits that he is currently agnostic about the question of whether or not the big bang event is a true *beginning* for our space-time.

[144] (Vilenkin [2006], p. 175 emphasis mine). Abhay Ashtekar ([2009], p. 9), a loop quantum cosmology proponent, acknowledged that the BGV does not rely on Einstein's field equations.

[145] In fact, the theorem is sometimes referred to as "the Ehlers-Geren-Sachs theorem of general relativity". (Clarkson, Coley, and O'Neill [2001], p. 063510-1).

[146] After all, is it not the case that both loop quantum cosmology (see Bojowald [2001], [2005]; and Ashtekar [2009]) and string-theoretic cosmology do away with initial singularities? Perhaps. Remember though that string cosmology is inconsistent with the CC-M since such approaches are incompatible with C&C's background space. Moreover, recall that loop quantum gravity implies the *N*-bound, and the *N*-bound implies that entropy cannot be unbounded from above (see sect.3). And besides, adding a loop quantum cosmology to the CC-M transmutes that model into something quite different. C&C explicitly reject loop quantum cosmology in their ([2004], p. 29. They reject it because they believe it "invokes special low-entropy conditions on some Cauchy surface…").

While the GSL does not hold for any and all horizons, it does hold for de Sitter horizons[149], "any future-infinite timelike worldline"[150], and "every state of the universe".[151] Moreover, given that the GSL holds for every state, its time-reverse will hold for every state.[152] The time-reverse GSL says "that for any past-infinite worldline $W_{past}$, the past horizon $H_{past} = \partial I^+(W_{past})$ cannot *increase* as time passes…"[153]

Now, what Wall shows is the following equivalence:

(7): The GSL is true, just in case, given that there is some null surface $F$ according to which the generalized entropy is diminishing on $F$ at an arbitrary point, $F$ is not a horizon.[154]

But (7) implies:

(8): It is not the case that there is an infinite (toward the future) worldline $W_{fut}$, which relates to $F$ in such a way that $F$ is—for the relevant observer—a future horizon.[155]

Therefore, by Wall's theorem 3, some null surfaces cannot be indefinitely extended.[156] This conclusion can be tied to two assumptions (*viz.*, that the GSL holds, and that global hyperbolicity holds[157]) and then used to show that the relevant space-time (for which the assumptions hold, and

---

[147] It seems that C&C go in for a generalized second law. In their discussion of black hole entropy and Hawking radiation, they stated that "one can prove [69], [70], [71], [72] certain versions of the Generalized Second Law, which guarantees that the radiation itself, free to escape to infinity, does have a larger entropy than the original black hole." (Carroll and Chen [2004], p. 18)

[148] Or, with respect to any causal horizon, the sum of the horizon entropy, plus the field entropy external to any such horizon will necessarily increase as time marches forward (Wall, [2012], p. 104049-1). Interestingly, the GSL implies that the thermodynamic behavior of certain open systems (*e.g.*, a causal horizon's exterior) is akin to that of certain closed systems (ibid.).

I should add here that Wall is chiefly concerned with the *fine-grained* GSL defined in (Wall ([2013], p. 6, *cf.* p. 10)). The fine-grained version of the GSL requires a "fine-grained…definition of the state…used to compute…entropy." ([ibid.], p. 10) This means that the state one uses for computational purposes represents "the complete information about a state", and not just "information available to an observer". ([ibid.], p. 6) What I say in the main text above is true for the fine-grained GSL. So understand all subsequent reference to the GSL as reference to the fine-grained GSL.

[149] (Wall [2012], p. 104049-1). Davies ([1984]), and Davies *et. al.* ([1986]) argued that a GSL applies to de Sitter space, though *cf.* Davis, Davies, and Lineweaver ([2003]).

[150] (Wall [2013], p. 9).

[151] (Wall [2013], p. 10).

[152] (Wall [2013], p. 10, and see also [2009]).

[153] (Wall [2013], p. 10) emphasis in the original.

[154] Paraphrased from (Wall [2013], p. 18).

[155] (Wall [2013]).

[156] Theorem 3 is stated and proven in (Wall [2013], p. 19).

[157] With respect to space-times like our metagalaxy, Garrett DeWeese wrote, "…all standard models of the Big Bang are globally hyperbolic" (DeWeese [2004], p. 82). Moreover, global hyperbolicity trivially holds for de Sitter space-time, since that space-time features a global Cauchy hypersurface. See (Geroch [1970], and Hawking and Ellis [1973], pp. 209-10; *cf.* p. 263; *cf.* Bernal and Sánchez [2003]).

for which null surfaces cannot be indefinitely extended) is future-null-geodesically incomplete "because there is a singularity somewhere on [*F*]."[158]

A similar result can be proven given the time-reversed GSL. This means that one can show that the relevant space-time is past-null-geodesically incomplete.[159] Wall explicitly notes how his results can be understood within a quantum context ([2013], p. 20; pp. 32-37) and correctly observes that he has secured something like a quantum analog of Penrose's ([1965]) singularity theorem.

With respect to an application of Wall's theorem to our FLRW metagalaxy, he stated:

> Putting all these considerations together, if the GSL is a valid law of nature, it strongly suggests that either the universe had a finite beginning in time, or else it is spatially finite and the arrow of time was reversed previous to the Big Bang. In the latter case, it could still be said that the universe had a beginning in a thermodynamic sense, because both branches of the cosmology would be to the thermodynamic future of the Big Bang.[160]

Of course, the CC-M posits an eternally inflating FLRW sub-model of our metagalaxy. Thus, the reversed arrow of time idea cannot be added to that sub-model.[161] We can conclude then, that Wall provides us with yet another reason for why we ought to believe that our metagalaxy is past-null geodesically incomplete.[162] This, I believe, serves as a significant defeater for the claim that our metagalaxy nucleated by means of the FGG mechanism from a background de Sitter space.

### 5.4 Fluctuation

The means by which our metagalaxy came forth out of a background space need not have involved a quantum tunneling process like the one recommended by FGG. In fact, C&C's original paper ([2004]) did not use the FGG mechanism. Rather, it urged that a suitable proto-inflationary patch could have—*via* the harmonic oscillation of a potential—fluctuated into existence from the background de Sitter space. But C&C believe that the probability that the relevant patch should fluctuate into existence by means of the recommended process is incredibly small. And that this patch should spark the process of eternal inflation is also regarded as incredibly improbable.[163] In fact, the probability is so small that C&C describe it as possibly "the smallest positive number in the history of physics…"[164] C&C can acknowledge wholeheartedly such a small probability without fear or trepidation because their model is very much a "wait and see" model (*cf.* McInnes [2007] p. 8). Because the background space-time is eternal, and geodesically complete, fluctuations of just the right sort will inevitably occur, *a fortiori*, they will occur an infinite amount of times. On this "wait and see" feature of the model, C&C stated:

> The total spacetime volume of the de Sitter phase will continue to increase, just as in eternal inflation. The total spacetime volume of the de Sitter phase is therefore infinite, and the transition into our proto-inflationary universe is guaranteed eventually to occur. *Indeed, it will eventually occur an infinite number of times*.[165]

---

[158] (Wall [2013], p. 19]). And see the proof for this in ([ibid.], pp. 19-20).

[159] (Wall [2013], p. 20).

[160] (Wall [2013], p. 27).

[161] Plus C&C rejected that picture when they rejected the Aguirre-Gratton model. See (Carroll and Chen [2004], p. 29).

[162] You might maintain that C&C need not appropriate the FGG proposal. There are, after all, suggested improvements of the tunneling story told there. Why then cannot C&C simply side-step the objections in this section by appropriating one of these ameliorations. The problem is that improvements like the one in (Fischler, Morgan and Polchinski [1990]) fail if our bubble metagalaxy features an initial singularity. That is why they diligently seek to rub initial singularities out (see [1990], pp. 4046-7).

[163] (Carroll and Chen [2004], p. 25). There is also the separate question of how likely it is that our metagalaxy's large scale structure is due to some prior inflationary era. Carroll and Tam address this question to some degree in their ([2010]).

[164] (Carroll and Chen [2004], p. 26. n. 4).

The point bears repeating. Because the de Sitter vacuum is both unstable and eternal, anything that can physically occur, will occur, and it will occur an infinite amount of times.[166]

One can see how the infinities are in some sense compounded on the CC-M once one realizes that the mechanism for producing the large scale structure of our metagalaxy is eternal inflation. According to Alan Guth, on such a sub-model, "anything that can happen will happen: in fact, it will happen an infinite number of times."[167] The latter implication of eternal inflation is relevant since—you will remember—the means by which entropy increases without bound is through the birth of metagalaxies. Because our metagalaxy will evolve into a de Sitter space, it will eventually start to behave like the background Universe, and spawn proto-inflationary patches which eternally inflate into even more metagalaxies. But you see, Guth's point is that eternal inflation also implies the inevitable birth of other metagalaxies without the extra thesis that our metagalaxy is an asymptotically de Sitter space-time. For on eternal inflation, certain regions of space never stop inflating. Some of these inflating regions will give birth to other universes in which physical constants and parameters may vary.[168]

So the background universe yields an infinite amount of metagalaxies, and an infinite amount of these will, through eternal inflation, yield an infinite amount of metagalaxies as well. What's the problem? The problem is that this wreaks havoc on probability judgments. If your sample space is infinite, it does not appear possible to have a well-defined probability measure to underwrite your probability and likelihood judgments. This problem of infinities and probabilities in eternal inflation-based cosmologies is well-known.[169] However, it is also well-known that there is no current satisfactory solution to the problem. In fact, Steinhardt noted that "[m]any remain hopeful even though they have been wrestling with this issue for the past 25 years *and have yet to come up with a plausible solution*."[170]

---

[165] (Carroll and Chen [2004], p. 27 emphasis mine. *Cf.* p. 23)

[166] This is a general property of the quantum vacuum. See (Redhead [1995a], [1995b]; and Summers [2012]).

[167] (Guth [2004], p. 49). Steinhardt ([2011], p. 43) stated, "[t]he truth is that quantum physics rules inflation, and anything that can happen will happen."

[168] See on this (Linde [2004], pp. 431-432; and Steinhardt [2011], p. 42).

[169] See (Page [2008], p. 063536-1) and the literature cited there.

[170] (Steinhardt [2011], p. 42 emphasis mine). Guth ([2004], p. 50) concluded similarly, "…we still do not have a compelling argument from first principles that determines how probabilities should be calculated." Elsewhere, Steinhardt and Turok ([2005], p. 44) remarked:
> "What is the probability distribution? In models such as eternal inflation, the relative likelihood of our being in one region or another is ill-defined since there is no unique time slicing and, therefore, no unique way of assessing the number of regions or their Notice that my criticism here would run even if C&C dispensed with eternal inflating sub-models. The problem of infinities appears when theorizing about ultra-large-scale structure (*i.e.*, the Universe). The problem is compounded when eternal inflating sub-models of metagalaxies such as our own are added in. I conclude then, that while C&C's original paper does not invoke the FGG mechanism (despite judgments from Carroll to the contrary), a heretofore unresolved theoretical problem remains, the problem of infinity and likelihood.[171]

## 6  Causation and the CC-M

Ignoring the CC-M's incompleteness, I have argued that it still fails to provide an adequate scientific explanation of our initial low-entropy state, since two claims essential to the CC-M (claims (2) and (3)) are false. By appeal to an argument for the well-foundedness of the causal relation, I will now argue that claim (1) is false as well.

### 6.1 Preliminaries

For the purposes of the main argument in this section, I will assume that purely contingent facts are proper *relata* for obtaining causal relations. Such facts are particular kinds of concrete states of affairs involving contingent substances or substrates exemplifying properties or standing in relations.[172] The nexuses of exemplification within purely contingent facts tie together *contingent* substrates (and only contingent substrates) with respective properties.[173] The relations within purely contingent facts relate contingent substrates and only contingent substrates to one another. Moreover, the improper parts of all purely contingent facts must themselves be contingent. An improper part of an object *o* just is *o*.[174] Thus; no purely contingent fact exists in every single world.[175]

I allow for complex purely contingent facts. Call such complexes higher-order purely contingent facts. Such higher-order purely contingent facts are themselves purely contingent in the sense that only contingent substrates fuse together and help comprise the respective high-order sums. Higher-order purely contingent facts are therefore purely contingent. This holds only if the following principle is true:

[Interpretation: ∇x: *x* is purely contingent (where what it means to be purely contingent is to only have *substrates that are contingent* as proper parts or constituents, though the entity in question may have properties or relations as

---

volumes. Brave souls have begun to head down this path, but it seems likely to us to drag a beautiful science towards the darkest depths of metaphysics."

[171] Even if a measure were found, one must still overcome the Boltzmann Brains problem. See (Page [2008]), and the literature cited therein.

[172] Michael Tooley believes that states of affairs are causal *relata*. See (Tooley [2003], p. 408).

[173] Reminiscent of Chisholm's understanding of events in his ([1990], p. 419 see definition D11). See also (Koons [2000], pp. 31-43).

[174] (Simons [1987], p. 11).

[175] This is one reason why [Cicero is Cicero] is not a purely contingent fact, for I consider that fact to be such that it exists at all worlds since it is true *in* all worlds and I (waiving in the direction of Thomas Crisp, and Timothy Williamson) am skeptical of a *truth-in/truth-at* distinction. I am also assuming the falsity of necessitism (the thesis that necessarily, every entity is necessarily some entity). I do find Williamson's ([2013]) defense of necessitism convincing, and so I merely assume contingentism (the negation of necessitism) here for the purposes of deliberation, since most philosophers seem to be contingentists.

constituents); Cxy: *x* is substrate-composed solely of *y* (where what it means for *x* to be substrate-composed solely of *y*, is for *x* to have *y* as its only proper part that is a substrate, though *x* may also have as proper parts or constituents properties or relations); Domain: unrestricted]

(**Purely Contingent Parts Principle, PCPP**): ■∀x(∃ys((Cxys & ∇ys) ⊃ ∇x))

That is to say, necessarily, for any entity *x*, if there are some *ys*, such that *x* is substrate-composed solely of the *ys*, and the *ys* are purely contingent, then *x* is purely contingent. The above principle makes use of "plural referring expressions" or plural quantification, but as Peter van Inwagen has said, that idea "has sufficient currency" contemporarily.[176, 177] PCPP seems to me to be highly intuitive. I do not have an argument for it, and so I should perhaps adopt a generally good dialectical strategy at this point and place a defeasibility operator 'Đ' in front of PCPP:

(**PCPP\***): Đ■[∀x(∃ys((Cxys & ∇ys) ⊃ ∇x))]

That is to say, normally, necessarily, for any entity *x*, if there are some *ys*, such that *x* is substrate-composed solely of the *ys*, and the *ys* are purely contingent, then *x* is purely contingent.[178]

I have argued in more than one place, that all purely contingent facts have causes.[179] I will not revisit my reasoning for such a conclusion here. I will simply assume the universality of causation, and proceed in demonstrating that the causal relation is well-founded with respect to purely contingent facts.

6.2 The Well-Foundedness of Causation

In his very important work, *Realism Regained: An Exact Theory of Causation, Teleology, and the Mind* ([2000]), Robert Koons proffered a very interesting argument for the well-foundedness of the causal relation (see p. 113) which depended upon the universality of causation. A close cousin of that argument proceeds as follows.

All purely contingent facts have causes, and there is a purely contingent fact *m* that is the sum of all purely contingent facts. If all purely contingent facts have causes, then *m* has a cause, call it *c*. Moreover, if for any obtaining causal relation which composes *m*, the cause in such a relation is preempted by *c*, then it is not the case that there is a complex purely contingent fact *m* that is the sum of all purely contingent facts, since every would-be cause would fail to actually bring about the relevant effect. Now, if *c* causes *m*, then either, for any obtaining causal relation which composes *m*, the cause in such a relation will be preempted by *c*, or else *c* causes *m* by indirectly (through the transitivity of causation) causing all of its constituent purely contingent facts by being the *initial* cause of *m*'s earliest obtaining purely contingent fact or facts. However, if *m* were an infinitely long causal chain whose links involved only purely contingent facts, then it would be false that *c* causes *m* by indirectly (through the transitivity of causation) causing all of its *initial* constituent purely contingent facts. But now, assume that *m* is an infinitely long causal chain whose links involve only purely contingent facts. It will now follow that the proposition <*c* causes *m* by indirectly (through the transitivity of causation) causing all of its constituent purely contingent facts by being the *initial* cause of *m*'s initial obtaining purely contingent fact or facts> is false. Additionally, it follows that if *c* causes *m*, then for any obtaining causal relation which composes *m*, the cause in such a relation is preempted by *c*. And so if *m* has a cause *c*, then it is not the case that there is a complex purely contingent fact *m* that is the sum of all purely contingent facts. But the universality of causation entails that there is a cause *c* of *m*. Therefore, it is not the case that there is a complex purely contingent fact *m* that is the sum of all purely contingent facts. But now we have an absurdity: there is a purely contingent fact *m* that is the sum of all purely contingent facts, and it is not the case that there is a purely contingent fact *m* that is the sum of all purely contingent facts. We can now safely conclude that it is not the case that *m* is an infinitely long causal chain whose links involve only purely contingent facts.

---

[176] (van Inwagen [1990], p. 23).
[177] For more on plural quantification, see (Uzquiano [2003]; van Inwagen [1990], pp. 23-32).
[178] On defeasibility reasoning see (Koons [2013]).
[179] (Weaver [2012], [forthcoming]).

Why would one maintain that:

(9): If *m* has a cause *c*, then either for any obtaining causal relation which composes *m*, the cause in such a relation is preempted by *c*, or else *c* causes *m* by indirectly (through the transitivity of causation) causing all of its constituent purely contingent facts by being the *initial* cause of *m*'s earliest obtaining purely contingent fact or facts.

Why could we not uphold the claim that *c* causes *m* by causing—*via* overdetermination—all of *m*'s constituent purely contingent facts? Koons ([2000], p. 113) fails to (at least) *explicitly* defend the supposition that we are dealing here with a case of preemption instead of a case of *symmetric overdetermination*, or even *joint causation*.

There are several ways to supplement Koons's argumentation. We might follow Martin Bunzl who argued that symmetric overdetermination is impossible (see Bunzl [1979]), though he still admitted that there is *explanatoral* overdetermination see p. 145), but Bunzl's reasoning requires a specific analysis of events, particularly the analysis of Donald Davidson ([1967]).[180] Davidson's analysis includes identity or individuation conditions for events (Davidson [1969], [2001], p. 179), where some event $e_1$ is identical to some other event $e_2$, just in case, for any event *x*, *x* directly causes $e_1$ just in case, *x* directly causes $e_2$, and for any event *x*, $e_1$ directly causes *x* just in case, $e_2$ directly causes *x* (see [ibid.]; and the discussion in Simons [2003], p. 374). This view of the identity conditions of events is flawed, for as Myles Brand ([1977], p. 332) pointed out, it equates all events which do not have direct causes, and which do not directly cause other events. It is an undesirable consequence that all ineffectual events are identical.[181] So we should abandon Davidson's analysis of events because of its view of the identity conditions of causal *relata*.

There is a different path for defending a related but weaker thesis: With respect to the actual world, there are no cases of overdetermination. One might have good reasons for believing in explanatory exclusion, and causal closure.[182] Some believe that if those two dogmas hold, then there are no actual cases of overdetermination. Wim De Muijnck goes further, "[m]etaphysically speaking, no such thing as overdetermination seems possible; this is a consequence…of causal closure and explanatory exclusion".[183] I, however, find causal closure to be objectionable, and so it is best to look for proper substantiation of (9) elsewhere.

---

[180] See (Bunzl [1979], p. 145, and p. 150).
[181] I do believe that some events do not have causes (*e.g.*, the event or state of affairs of God's causing the cosmos).

Some philosophers have suggested that symmetric overdetermination is improbable.[184] One response to this charge which draws from (Schaffer [2003], pp. 27-29; and Paul [2007]) is that "quantitative"[185], and/or "constitutive overdetermination"[186] is prevalent. If one is a non-reductionist about the structure of material objects, then a great many cases of macroscopic causation will involve Paul's constitutive overdetermination, in that the parts which compose such wholes (any macro-level material entity) will contribute to causally producing that which the macro-level entity (the whole) brings about (*cf.* Paul[187] [2007], pp. 276-277). Schaffer recommends that one fend off the objection that parts of the causally efficacious object are not metaphysically distinct enough to breed overdetermination, by noting that the parts are "*nomologically* correlated" while still "*metaphysically* distinct".[188] I agree with Schaffer's response.

Let me recommend a different strategy. Give attention to a standard explication of symmetric overdetermination as articulated by L.A. Paul ([2007], pp. 269-270 emphasis mine):

> In contemporary discussions of causation, standard cases of symmetric causal overdetermination are defined (roughly) as cases involving multiple distinct causes of an effect where the causation is neither joint, additive, nor preemptive (*and it is assumed the overdetermining causes do not cause each other*)…Each cause makes exactly the same causal contribution as the other causes to the effect (so the causal overdetermination is *symmetric*); each cause without the others is sufficient for the effect; and for each cause the causal process from cause to effect is not interrupted.[189]

---

[182] Kim defined causal closure as the thesis that "…*any physical event that has a cause at time t has a physical cause at t.*" (Kim [1989b], p. 43 emphasis in the original). He says that explanatory exclusion is the principle that "[n]o event can be given more than one *complete* and *independent* explanation." (Kim [1989a], p. 79 emphasis in the original).

[183] De Muijnck ([2003], p. 65).

[184] You get a hint of this in (Kim [1989a], p. 86; Schaffer cites Kim [1998] along these lines). *Cf.* the discussion in (Schaffer [2003], pp. 27-9 and see footnote 6 on p. 41).

[185] Mackie's term, from Mackie ([1974], p. 43); *cf.* De Muijnck ([2003], pp. 65-6); Schaffer ([2003], p. 28). Schaffer tells us that, "…quantitative overdetermination occurs whenever the cause is decomposable into distinct and independently sufficient parts". Schaffer ([2003], p. 28).

[186] Paul's term from (Paul [2007]).

[187] I should point out that Paul believes that the consequence of there being such prevalent constitutive overdetermination is "mysterious and problematic". Paul ([2007], p. 277). She attempts to rid the world of such prevalence by arguing that fundamental causal *relata* are property instances shared by overlapping entities involved in obtaining causal relations. See (Paul [2007], p. 282).

[188] (Schaffer [2003], p. 42. n. 9 emphasis in the original). The objection is tied to (Kim [1989a]). Some interpret Kim as suggesting that overdetermination just doesn't involve the causation of an event by an object and the parts which compose it. See (Sider [2003], p. 719. n. 2).

[189] See also (Sider [2003], p. 719. n. 2).

Now pick out an arbitrary obtaining causal relation which composes *m*. Label the cause *D*, and the effect *E*. If we understand symmetric overdetermination in the way Paul recommends, we cannot say that *c* is a direct overdetermining cause with *D* of *E*, precisely because *c* is a cause of *D* in that it is the cause of *m*. Perhaps a charitable reading of (Koons [2000], p. 113) would suggest that by his admission that (in the case as I have adjusted it) *c* is "causally prior" to *D* in that *c* is also the cause of *D*, *c* and *D* cannot be overdetermining causes of *E*. On the charitable reading, my attempted improvement on (Koons [2000], p. 113) *per* this pericope only involves further elaboration upon just how causal priority precludes a symmetric overdetermination reading of the relevant case.

One might think that the charitable reading does not help preclude a joint causation understanding of the case of concern. But this is not right. We were supposing in the Koons-style reasoning above that *D* is a sufficient cause of *E*. So *D* is enough to bring about *E*, and since we are not dealing here with symmetric overdetermination, the relationship between *c*, *D*, and *E* must be understood in terms of asymmetric overdetermination, *i.e.*, preemption.

The above Koons-style argument will generalize to any infinitely long causal chain composed of obtaining causal relations which feature only purely contingent facts.

6.3 Well-Foundedness and the CC-M

The background space of the CC-M contains two actual infinities on each side of the "initial" Cauchy hypersurface.[190] Each infinity spawns a sea of infinite metagalaxies. There therefore seems to be two senses in which the CC-M implies the existence of an infinitely long causal chain.

First, the actual infinities in the background space involve an ongoing infinite process of cosmic evolution *via* the evolution of the de Sitter space-time. We can choose a random arbitrarily large surface of the space-time, sum it up and judge that such a surface will serve as a causal dependency base for some subsequent exceedingly large surface of the same space-time. That surface will also serve as a causal dependency base for some succeeding surface and so on *ad infinitum*. Thus, the CC-M implies that there is an infinitely long causal dependency chain.[191] Such dependency will suffice as insurance for respective obtaining causal relations, and so the CC-M likewise implies that there is an infinitely long causal chain.

The other sense in which the CC-M may imply an infinitely long chain of obtaining causal relations is connected to metagalaxy nucleation. Carroll seems to regard the pinching off of a metagalaxy from the Universe as a causal process. Thus, if that process really does obtain infinitely many times, then we can sum up an arbitrary metagalaxy with the first proper parts of the nucleation process including the pinching off phenomena. That sum will serve as a causal dependency base for the birth of a metagalaxy. But because there is bound to be somewhere an infinitely long chain of "just so" birthing (*i.e.,* a chain in which asymptotically de Sitter space-times give birth to more and more asymptotically de Sitter space-times), daughters of space-time birthing processes can sum up with their own birthing processes and thereby serve as causal dependency bases for their newborns. The CC-M in this way implies an infinitely long causal chain.

---

[190] That is, according to the standard way of understanding de Sitter space-times.
[191] In the discussion above, I assume a simple counterfactual analysis of causation, not unlike the one defended in (Lewis [1973]) or the later more sophisticated analysis in (Lewis [2004]). My point could easily be adjusted so as to accommodate other, (perhaps) more plausible accounts of causation.

If at least one of the above recommendations of connecting the CC-M to the implication that there is an infinitely long causal chain is sound, then the Koons-style argument for well-foundedness will run, and the CC-M will come out false.

## 7 Conclusion

The CC-M is admittedly, though still woefully, incomplete. This incompleteness transfers to its proposed scientific explanation of our initial low-entropy state. Even if we grant that the model and explanation are in some sense complete, all of its essential claims are false. We should therefore refrain from looking to the CC-M for a dynamical explanation of the arrow of time.

**Acknowledgements:** I would like to thank David Albert, Tom Banks, Barry Loewer, Don N. Page, Quayshawn Spencer, and Aron C. Wall for helpful comments on the paper. Special thanks to Anthony Aguirre, Robin Collins, and Matthew Johnson for valuable conversations about some of the arguments presented here. And let me express gratitude and thanks to Sean Carroll for kindly and patiently answering several of my questions about the model during the UC Santa Cruz Summer Institute for the Philosophy of Cosmology in the summer of 2013.

**Christopher Gregory Weaver**
**Teaching Assistant**
**PhD student**
**Philosophy Department, Rutgers University**

**Contact Info:**
**Rutgers, the State University of New Jersey**
**Department of Philosophy**
**Attn: Christopher G. Weaver**
**110 Somerset Street, 5th Floor**
**New Brunswick, NJ 08901-4800**
**Email:** christophergweaver [at] gmail [dot] com